\documentclass[a4paper]{jpconf}
\usepackage{graphicx}
\usepackage{bbold}
\usepackage{slashed}
\usepackage{amsmath}
\usepackage{amssymb}

\newcommand{\MPl}{M_{\rm Pl}}
\newcommand{\bea}{\begin{eqnarray}}
\newcommand{\eea}{\end{eqnarray}}
\newcommand{\be}{\begin{equation}}
\newcommand{\ee}{\end{equation}}
\newcommand{\tmop}[1]{\ensuremath{\operatorname{#1}}}

\begin{document}
\begin{flushleft} 
KCL-PH-TH/2020-{\bf 62}
\end{flushleft}

\title{Non-Hermitian Yukawa  interactions of fermions with axions: potential microscopic origin  and dynamical mass generation}

\author{Nick E. Mavromatos}

\address{King's College London, Department of Physics, Theoretical Particle Physics and Cosmology Group, Strand, London WC2R 2LS, UK}

\ead{nikolaos.mavromatos@kcl.ac.uk}

\begin{abstract}
In this mini review, we discuss some recent developments regarding properties of (quantum) field-theory models containing anti-Hermitian Yukawa interactions between pseudoscalar fields (axions) and Dirac (or Majorana) fermions. Specifically, after motivating physically such interactions, in the context of string-inspired low-energy effective field theories, involving right-handed neutrinos and axion fields, we proceed to discuss their formal consistency within the so-called Parity-Time-reversal(PT)-symmetry framework, as well as dynamical mass generation, induced by the Yukawa interactions, for both fermions and axions. The Yukawa couplings are assumed weak, given that  they are conjectured to have been generated by non-perturbative effects in the underlying microscopic string theory. The models under discussion contain, in addition to the Yukawa terms, also anti-Hermitian anomalous derivative couplings of the pseudoscalar fields to axial fermion currents, as well as interactions of the fermions with non-Hermitian axial backgrounds. We discuss the role of such additional couplings on the Yukawa-induced dynamically-generated masses. For the case where the fermions are right-handed neutrinos, we compare such masses with the radiative ones induced by both, the anti-Hermitian anomalous terms and the anti-Hermitian Yukawa interactions in phenomenologically relevant models.  
\end{abstract}

\section{Introduction \label{sec:intro}}

\medskip

The Parity-Time-reversal(PT)-symmetry framework~\cite{PTqm1,PTqm,PT,PT2} is an innovative approach to quantum theory, with a plethora of theoretical and experimental applications in various branches of physics (for a partial but indicative list of such applications, which are rapidly expanding, though, to embrace new phenomena, even as this review is being written, the reader can consult the mini review~\cite{PTappl}.) PT symmetry guarantees the self consistency of quantum mechanical models with non-Hermitian Hamiltonians, characterised by real energy eigenvalues. The reality of the energy eigenvalues, despite the lack of Hermiticity of the Hamiltonian, can be understood~\cite{most2,most3,most,maninnerPT,antilin} by means of the {\it antilinear} nature of PT symmetry. In fact, as argued in \cite{antilin},  PT symmetry constitutes only a special case of non-Hermitian Hamiltonians with real eigenvalues. If a quantum system is characterised by an {\it antilinear symmetry}, this is the most general condition that one can impose on a quantum theory for which one can have a time-independent inner product and a self-adjoint Hamiltonian  with real energy eigenvalues. For each of the above properties Hermiticity is only a sufficient condition but not a necessary one. Hermiticity is then a special case, in which the Hamiltonian of the system has both antilinearity and Hermiticity.\footnote{We also note that {\it spontaneous breaking} of PT symmetry, as for instance is the case in the 
(1+1)-dimensional quantum  mechanical lattice system studied in \cite{cherno2}, leads to the existence of an energy spectrum with complex branches.}

Extension of the methods of PT-symmetric non-Hermitian quantum-mechanics towards the formulation of non-Hermitian quantum field theories within the PT-symmetry framework is at present in its initial stages, but with very encouraging results~\cite{qft1a,qft1b,qft1c,qft1d,qft1e,qft2,qft3,qft4,AB,AMS,qft5}. There are studies in this framework associated with  Dirac fermion field theories~\cite{ptf},  neutrino models~\cite{neu1,neu2}, spontaneous breaking of global and local (gauge) symmetries in non-Hermitian field theories~\cite{hg1,hg2,hg3,hg4,hg5,hg5b,hg6}, discrete symmetries~\cite{ptdisc} and supersymmetry~\cite{ptsusy} in such models,  chiral magnetic effect in 
non-Hermitian fermionic systems~\cite{chernochiral}, (1+1)-dimensional time-like Liouville conformal field theories~\cite{liouv},  a discussion on the potential r\^ole of non-Hermitian Hamiltonians for the stability of the Higgs vacuum and other field theories of interest to particle physics~\cite{stab}, 
as well as studies of non-Abelian magnetic monopole solutions~\cite{ptmon}, and more generally complex Bogomol'nyi-Prasad-Sommerfield (BPS) solitons with real energies~\cite{bpssol}. 

All the above field-theoretic systems are relativistic, for which the reality of the energy eigenvalues can be understood by the extension of the quantum mechanical ideas of the existence of an underlying antilinear symmetry~\cite{most,maninnerPT,antilin} to quantum field theories with non-Hermitian Hamiltonians. The antilinear symmetry is uniquely identified with $\mathcal C$PT~\cite{antilin}, where $\mathcal C$ is an appropriate definition of the charge conjugation, which may differ from the standard definition of the Dirac conjugation operator~\cite{AMS}. Indeed, as shown in \cite{antilin}, requiring the existence of time-independent inner products and invariance under {\it complex} Lorentz transformations, forces the antilinear symmetry to be {\it uniquely} $\mathcal C$PT ~\cite{antilin}. In this way, the standard CPT theorem (with C denoting the standard Dirac charge-conjugation operator), which is based on locality, unitarity and Lorentz invariance of the corresponding field-theoretic Lagrangian densities, can be extended to appropriate field-theoretic systems with non-Hermitian Hamiltonians. In this latter $\mathcal C$PT-invariant framework, PT-symmetric systems are  charactrerised by  a separate invariance under charge conjugation $\mathcal C$~\cite{antilin}.

In the above examples, the non-Hermitian Hamiltonians have been assumed, without an attempt to provide microscopic explanations. In \cite{alex1,alex2,soto} we have discussed possible microscopic explanations for a particular kind of non-Hermitian interactions, that of {\it anti-Hermitian Yukawa} interactions between pseudoscalar (axion-like) and fermion fields, within the framework of certain string-inspired models~\cite{mp}. The underlying $\mathcal C$PT symmetry that characterises the model is responsible for the reality of the energy eigenvalues, according to the general arguments of \cite{most2,most3,most,maninnerPT,antilin}, mentioned above. 
Moreover, we have studied dynamical mass generation for the pseudoscalar and fermion fields in such systems, within a non-perturbative Schwinger-Dyson (SD) treatment. 
It is the purpose of this article, to briefly review these studies. 

The structure of the article is as follows: in the next section, \ref{sec:model}, 
we review the string-inspired model~\cite{mp} and explain how the non-Hermitian Yukawa interactions emerge, along with non-Hermitian anomalous couplings of the pseudoscalar (axion-like) fields. We note at this point that the axion-like particles in this model are associated with stringy excitations and are in general different from the QCD axion. In the literature such fields are sometimes called axion-like particles (ALPs), a terminology used in \cite{alex2,soto} where this review is partly based on. 
For notational brevity, though, in what follows, we shall refer to them simply as axions. We motivate the use of non-Hermitian Yukawa models (embedded in appropriately generalised-PT ($\mathcal C$PT) frameworks) as providers of alternative ways of  dynamical-mass generation for  {\it both} axions and fermions. In the string-inspired model in which the fermions are right-handed neutrinos, both these fields and the axions in their massive phase could provide candidates for dark matter. In  section \ref{sec:dynamical}, we discuss the SD mass generation for axions and fermions, in the absence of anomalous axion couplings. In section \ref{sec:anom}, we include such anomalous interactions of axions, and study their effects on the anti-Hermitian-Yukawa-interaction-induced mass generation.
We also include external non-Hermitian axial backgrounds, which are motivated within the context of the string-inspired model of section \ref{sec:model}. Conclusions and outlook are presented in section \ref{sec:concl}. 

\medskip

\section{String-inspired Models and a potential origin of non-Hermitian Interactions \label{sec:model}}

\medskip

In this section we shall motivate the origin of non-Hermitian interactions of axions in low-energy effective actions inspired from string theory. To this end, we shall first review some basic features of the spectrum of string theories.
We shall be dealing with perturbative effective actions of string theory~\cite{str1,str2,pol1,pol2}, restricting ourselves to at-most-quadratic order in space-time derivatives. This suffices when we discuss physics at energy scales much below the string mass scale, which serves as an ultraviolet (UV) momentum-cut-off scale of the point-like effective field theories stemming from strings~\cite{str1,str2}. 

\medskip

\subsection{The Bosonic massless gravitational mutliplet of strings and anomalies \label{sec:massmult}}

\medskip 

In superstring theory~\cite{str1,str2,pol1,pol2}, after compactification to four space-time dimensions, the bosonic ground state of the closed-string sector consists of {\it massless} fields in the so-called {\it gravitational multiplet}, which contains a spin-$0$ (scalar) dilaton $\Phi (x)$, 
a spin-2 traceless symmetric tensor field, $g_{\mu\nu}(x)$, which is uniquely identified as the  (3+1)-dimensional graviton, and a spin-1 antisymmetric tensor gauge field $B_{\mu\nu}(x)=-B_{\nu\mu}(x)$, known as the Kalb-Ramond (KR) field. In what follows, for brevity and concreteness, we shall set the four-dimensional dilaton field to a constant, $\Phi(x)=\Phi_0$. This will fix the string coupling 
\be\label{stringcoupl}
g_s=\exp(\Phi) = \exp(\Phi_0)~. 
\ee
There are always consistent solutions of the four-dimensional string theory with such a configuration, and this suffices for our purposes in this review.

There is a U(1) gauge symmetry of the closed-string (3+1)-dimensional target-space-time effective-field-theory action, associated with the KR $B$-field transformations 
\be\label{Bgauge}
B_{\mu\nu}(x) \, \rightarrow \, B_{\mu\nu}(x) + \partial_{[\mu}\theta_{\nu]}(x), \quad \mu,\nu =0, \dots 3, 
\quad \theta_\mu (x) \in \mathbb R, 
\ee
where Greek indices from now on denote space-time indices, taking on the values $0, \dots 3$, and 
the symbol $[\dots]$ denotes antisymmetrization of the respective indices.

The U(1) gauge symmetry of the closed-string sector implies that the corresponding effective action will be expressed only in terms of the field strength of the $B$-field:
\be\label{KRfs}
\mathcal H_{\mu\nu\rho}(x) = \partial_{[\mu} B_{\nu\rho]}(x).
\ee
This is subject to the following Bianchi identity 
\be\label{bianchi}
 \mathcal H_{[\nu\rho\sigma\, ;\, \mu]} = \partial_{[\mu}\mathcal H_{\nu\rho\sigma]} = 0~,
\ee
where, from now on, we omit the space-time-coordinate arguments of the fields ($x$) for brevity. The semicolon denotes covariant derivative with respect to the standard
Christoffel connection $\Gamma_{\,\,\mu\nu}^\alpha= \Gamma_{\,\,\nu\mu}^\alpha$ of the metric $g_{\mu\nu}$. It should be stressed that, due to the total antisymmetry of $\mathcal H$, the terms involving the gravitational Christoffel connection drop out from \eqref{bianchi}.

In superstring theory, anomaly cancellation requirements imply a modification of the KR field strength \eqref{KRfs} by 
appropriate gauge (Yang-Mills (Y)) and Lorentz (L) (gravitational) Chern-Simons terms (Green-Schwarz (GS) mechanism)~\cite{str2}
\begin{align}\label{csterms}
\mathbf{{\mathcal H}} &= \mathbf{d} \mathbf{B} + \frac{\alpha^\prime}{8\, \kappa} \, \Big(\Omega_{\rm 3L} - \Omega_{\rm 3Y}\Big),  \nonumber \\
\Omega_{\rm 3L} &= \omega^a_{\,\,c} \wedge \mathbf{d} \omega^c_{\,\,a}
+ \frac{2}{3}  \omega^a_{\,\,c} \wedge  \omega^c_{\,\,d} \wedge \omega^d_{\,\,a},
\quad \Omega_{\rm 3Y} = \mathbf{A} \wedge  \mathbf{d} \mathbf{A} + \mathbf{A} \wedge \mathbf{A} \wedge \mathbf{A},
\end{align}
where $\alpha^\prime = M_s^{-2}$, with $M_s$ the string mass scale, which is in general different from the 
four-dimensional Planck mass scale $M_P = 1.22 \times 10^{19}~{\rm GeV} \equiv \sqrt{8\,\pi}\, \kappa^{-1} $. 
For notational brevity, in \eqref{csterms} we used differential-form language~\cite{eguchi}, with $\mathbf d$ denoting the 
exterior-derivative one-form, $\mathbf d=dx^\mu \, \partial_\mu$,  and $\wedge$ the exterior (``wedge'') product among differential forms, such that  ${\mathbf f}^{(k)} \wedge {\mathbf g}^{(\ell)} = (-1)^{k\, \ell}\, {\mathbf g}^{(\ell)} \wedge {\mathbf f}^{(k)}$, where ${\mathbf f}^{(k)}$, and ${\mathbf g}^{(\ell)}$ are $k-$ and $\ell-$ forms, respectively. Above, $\mathbf{A}$ is the Yang-Mills gauge field one-form, and $\omega^a_{\,\,b}$ the spin-connection one-form (the Latin indices $a,b,c,d$ are (3+1)-dimensional tangent-space ({\it i.e}. Lorentz-group-SO(1,3)) indices, referring to the Minkowski manifold which is tangent to the space-time manifold at a coordinate point $x$).

The addition of the Chern-Simons terms in \eqref{csterms} leads to a modification of the Bianchi identity (\ref{bianchi}), which can now be written as~\cite{str2}
\begin{align}\label{modbianchi2}
& \varepsilon^{\mu\nu\rho\sigma}\, \mathcal H_{[\nu\rho\sigma\, ;\, \mu]} =  \varepsilon_{abc}^{\;\;\;\;\;\;\mu}\, {\mathcal H}^{abc}_{\;\;\;\;\;\; ;\mu} 
 =  \frac{\alpha^\prime}{32\, \kappa} \, \sqrt{-g}\, \Big(R_{\mu\nu\rho\sigma}\, \widetilde R^{\mu\nu\rho\sigma} -
\mathbf F_{\mu\nu}\, \widetilde{\mathbf F}^{\mu\nu}\Big) \nonumber \\ & \equiv \sqrt{-g}\, {\mathcal G}(\omega, \mathbf{A}) = \partial_\mu \, \mathcal K^\mu (\omega, \mathbf A),
\end{align}
where the right-hand side  denotes the mixed anomaly, due to chiral fermions in the theory circulating in the anomalous loop~\cite{alvarez,weinberg}, $g$ denotes the determinant of the metric tensor, $\mathbf{F} = \mathbf{d} \mathbf{A} + \mathbf{A} \wedge  \mathbf{A}$ is the two-form corresponding to the Yang-Mills field strength  (we use form notation for brevity here),  
$R_{\mu\nu\rho\sigma}$ is the Riemann space-time curvature tensor\footnote{Our conventions and definitions used throughout this work are: signature of metric $(+, -,-,- )$, Riemann Curvature tensor
$R^\lambda_{\,\,\,\,\mu \nu \sigma} = \partial_\nu \, \Gamma^\lambda_{\,\,\mu\sigma} + \Gamma^\rho_{\,\, \mu\sigma} \, \Gamma^\lambda_{\,\, \rho\nu} - (\nu \leftrightarrow \sigma)$, Ricci tensor $R_{\mu\nu} = R^\lambda_{\,\,\,\,\mu \lambda \nu}$, and Ricci scalar $R = R_{\mu\nu}g^{\mu\nu}$.} 
and 
\begin{equation}\label{leviC}
\varepsilon_{\mu\nu\rho\sigma} = \sqrt{-g}\,  \epsilon_{\mu\nu\rho\sigma}, \quad \varepsilon^{\mu\nu\rho\sigma} =\frac{{\rm sgn}(g)}{\sqrt{-g}}\,  \epsilon^{\mu\nu\rho\sigma},
\end{equation}
with $\epsilon^{0123} = +1$, {\emph etc.}, are the gravitationally covariant Levi-Civita tensor densities, totally antisymmetric in their indices.
The symbol
$\widetilde{(\dots)}$
over the curvature- or gauge-field-strength tensors denotes the corresponding dual, defined as
\begin{align}\label{duals}
\widetilde R_{\mu\nu\rho\sigma} = \frac{1}{2} \varepsilon_{\mu\nu\lambda\pi} R_{\,\,\,\,\,\,\,\rho\sigma}^{\lambda\pi}, \quad \widetilde{\mathbf F}_{\mu\nu} = \frac{1}{2} \varepsilon_{\mu\nu\rho\sigma}\, \mathbf F^{\rho\sigma}.
\end{align}

The non-zero quantity on the right hand side  of \eqref{modbianchi2} is the ``mixed (gauge and gravitational) quantum anomaly''~\cite{weinberg,alvarez}, which is known to be a total divergence of a function $\mathcal K^\mu$ (containing the Chern-Simons forms in \eqref{csterms}).

To lowest order in the string Regge slope $\alpha^\prime$, the  
(3+1)-dimensional effective action of the closed-string bosonic sector is then given by~\cite{str1,str2,pol1,pol2}:
\begin{align}\label{sea2}
S_B =-&\; \int d^{4}x\sqrt{-g}\Big( \dfrac{1}{2\kappa^{2}}\, R + \frac{1}{6}\, {\mathcal H}_{\lambda\mu\nu}\, {\mathcal H}^{\lambda\mu\nu} + \dots \Big).
\end{align}
The KR field strength terms ${\mathcal H}^2$ in (\ref{sea2}) can be absorbed~\cite{str2} (up to an irrelevant total divergence) into a contorted generalised curvature
$\overline R (\overline \Gamma)$, with a ``torsional connection''~\cite{hehl} $\overline \Gamma$, corresponding to a contorsion tensor proportional to the totally antisymmetric ${\mathcal H}_{\mu\nu}^\rho$ field strength,
\begin{align}\label{torcon}
{\overline \Gamma}_{\mu\nu}^{\rho} = \Gamma_{\mu\nu}^\rho + \frac{\kappa}{\sqrt{3}}\, {\mathcal H}_{\mu\nu}^\rho  \ne {\overline \Gamma}_{\nu\mu}^{\rho}~,
\end{align}
where $\Gamma_{\mu\nu}^\rho = \Gamma_{\nu\mu}^\rho$ is the torsion-free (Riemannian) Christoffel symbol.

The reader should notice that the modifications (\ref{csterms}) and the right-hand-side of  (\ref{modbianchi2}) contain the {\it torsion-free} spin connection. In fact, it can be shown~\cite{hull,mavindex} that any potential contributions from the (totally-antisymmetric) torsion $\mathbf{H}$ three-form in the anomaly equation can be removed by adding to the string effective action appropriate counterterms order by order in perturbation theory.

Since the anomaly ${\mathcal G}(\omega, \mathbf{A})$ is an exact one loop result, one may implement the Bianchi identity (\ref{modbianchi2}) as a $\delta$-functional constraint in the quantum path integral of the action (\ref{sea2}) over the fields ${\mathcal H}$, $\mathbf{A}$, and $g_{\mu\nu}$, and express the latter in terms of a Lagrange multiplier (pseudoscalar) field~\cite{kaloper,witten,boss,anomalies} $b(x)/\sqrt{3}$ (where the normalisation factor $\sqrt{3}$ is inserted so that the field $b(x)$ will acquire a canonical kinetic term, as we shall see below) :
\begin{align}\label{delta}
&\Pi_{x}\, \delta\Big(\varepsilon^{\mu\nu\rho\sigma} \, {{\mathcal H}_{\nu\rho\sigma}(x)}_{; \mu} - {\mathcal G}(\omega, \mathbf{A}) \Big)
\Rightarrow  \nonumber \\ &\int {\mathcal D}b \, \exp\Big[i \, \,\int d^4x \sqrt{-g}\, \frac{1}{\sqrt{3}}\, b(x) \Big(\varepsilon^{\mu\nu\rho\sigma }\, {{\mathcal H}_{\nu\rho\sigma}(x)}_{; \mu} - {\mathcal G}(\omega, \mathbf{A}) \Big) \Big] \nonumber \\
&= \int {\mathcal D}b \, \exp\Big[-i \,\int d^4x \sqrt{-g}\, \Big( \partial ^\mu b(x) \, \frac{1}{\sqrt{3}} \, \epsilon_{\mu\nu\rho\sigma} \,{\mathcal H}^{\nu\rho\sigma}  + \frac{b(x)}{\sqrt{3}}\, {\mathcal G}(\omega, \mathbf{A}) \Big)\Big]
\end{align}
where  the second equality has been obtained by partial integration, upon assuming that the KR field strength dies out at spatial infinity. Inserting (\ref{delta})
into the ({\it Euclidean}) path integral with respect to the action (\ref{sea2}), and integrating over the ${\mathcal H}$ field, one obtains~\cite{kaloper,witten} an effective action in terms of the anomaly and a
dynamical, {\it massless}, KR-axion field $b(x)$, with canonically-normalised kinetic terms. 

\subsection{Ambiguities in the KR-axion effective action and non-Hermitian interactions \label{sec:ambig}}

\medskip 

There is a known ambiguity~\cite{strom} 
in analytically continuing the resulting effective action back to Minkowski space time, which stems from the following fact: the $\mathcal H$-field-strength Euclidean path integration results in the presence of a $b(x)$-dependent, quadratic in space-time derivatives, term in the bosonic (B) Euclidean (E) string effective action of the form: 
\be\label{square}
S_{\rm eff \, B}^{\rm (E)} \ni \int d^4 x \, \sqrt{g^{\rm (E)}} \frac{1}{12} \varepsilon_{\mu\nu\rho\lambda}^{\rm (E)} \, 
 \varepsilon^{\mu\nu\rho\sigma\, \rm (E)} \, \partial^\lambda b \, \partial_\sigma b.
\ee
where the covariant Levi-Civita tensor density $\varepsilon_{\mu\nu\rho\lambda}^{\rm (E)} $ has been defined in 
\eqref{leviC}, but here the notation (E) indicates that this quantity is evaluated in a Euclidean-signature metric (with the convention (+,+,+,+)). 
The ambiguity concerns the stage at which we analytically continue the quantity \eqref{square} back to Minkowski space-time. 

If we {\it first} use the following property of the Levi-Civita tensor density in four space-time dimensions with Euclidean metric (``\textbf{Scheme I'}'):
\be\label{propLC}
\varepsilon_{\mu\nu\rho\lambda}^{\rm (E)} \, 
 \varepsilon^{\mu\nu\rho\sigma\, \rm (E)} = + 6 \, \delta_\lambda^{\, \sigma}~, 
 \ee
where $\delta_\lambda^{\, \sigma}$ denotes the Kronecker delta, and {\it then} analytically continue to Minkowski space time, we obtain a real effective action for the dynamical field $b(x)$ (which plays the r\^ole of the KR gravitational axion):~\cite{kaloper}
\begin{align}\label{sea3}
S^{\rm eff (I)}_{\rm B} =&\; \int d^{4}x\sqrt{-g}\Big[ -\dfrac{1}{2\kappa^{2}}\, R + \frac{1}{2}\, \partial_\mu b \, \partial^\mu b +  \sqrt{\frac{2}{3}} \, \frac{\alpha^\prime}{96\, \kappa} \, b(x) \, \Big(R_{\mu\nu\rho\sigma}\, \widetilde R^{\mu\nu\rho\sigma} - \mathbf F_{\mu\nu}\, \widetilde{\mathbf F}^{\mu\nu}\Big) + \dots \Big],
\end{align}
where the dots $\dots$ denote gauge, as well as higher derivative, terms appearing in the string effective action, that we ignore for our discussion here.  In this construction the KR axion appears as a {\it standard pseudoscalar field}, with a canonically-normalised kinetic term with the {\it correct sign} relative to the space-time curvature (Eisntein-Hilbert)  terms in \eqref{sea3}. 
We also observe that, in view of the anomaly, the KR axion field couples to the gravitational and gauge fields. This latter interaction is P and T violating, and hence in view of the overall CPT invariance of the relativistic, local and unitary (quantum) field theory \eqref{sea3}, also CP violating (we remind the reader that C denotes the standard Dirac charge-conjugation operator).
 
On the other hand, if one {\it first} analytically continues \eqref{square} to a Minkowski-signature space time, and {\it then} 
uses the Minkowski version of \eqref{propLC} (``\textbf{Scheme II}''):
\be\label{propLCM}
\varepsilon_{\mu\nu\rho\lambda} \, 
 \varepsilon^{\mu\nu\rho\sigma} = - 6 \, \delta_\lambda^{\, \sigma}~, 
 \ee
where the minus sign on the right-hand side is due to the dependence of the contravariant Levi-Civita tensor density  \eqref{leviC}
on the (Minkowski) signature of the metric tensor, then, one obtains an effective action in which the kinetic terms of the $b$ field 
have the {\it wrong} sign relative to the space-time-curvature terms in  the effective action, and thus the KR axion would behave like a {\it ghost} field:
\begin{align}\label{sea3Mghost}
S^{\rm eff (II)}_{\rm B} =&\; \int d^{4}x\sqrt{-g}\Big[ -\dfrac{1}{2\kappa^{2}}\, R - \frac{1}{2}\, \partial_\mu b \, \partial^\mu b +  \sqrt{\frac{2}{3}} \, \frac{\alpha^\prime}{96\, \kappa} \, b(x) \,  \Big(R_{\mu\nu\rho\sigma}\, \widetilde R^{\mu\nu\rho\sigma} - \mathbf F_{\mu\nu}\, \widetilde{\mathbf F}^{\mu\nu}\Big) + \dots \Big],
\end{align}
Naively, one would ignore this second construction, because of the common perception that a ghost axion field does not carry any physical significance. However, in view of the PT-symmetric framework~\cite{PTqm1,PTqm,PT,PT2,most}
and its field-theory extensions, one should reconsider this point of view. Indeed, by viewing the $b$ field in \eqref{sea3Mghost}
as purely imaginary,
\be\label{btoib}
b(x) \to i\, b(x), \quad b(x) \in \mathbb R,  
\ee
one may write the corresponding effective action as
\begin{align}\label{sea3PT}
S^{\rm eff \, (II)}_{\rm B} =&\; \int d^{4}x\sqrt{-g}\Big[ -\dfrac{1}{2\kappa^{2}}\, R + \frac{1}{2}\, \partial_\mu b \, \partial^\mu b +  i\, \sqrt{\frac{2}{3}} \, \frac{\alpha^\prime}{96\, \kappa} \, b(x) \, \Big(R_{\mu\nu\rho\sigma}\, \widetilde R^{\mu\nu\rho\sigma} - \mathbf F_{\mu\nu}\, \widetilde{\mathbf F}^{\mu\nu}\Big) + \dots \Big],
\end{align}
which has a canonical kinetic term for the (redefined) axion field $b$, which no longer behaves as a ghost, but it is now characterised by non-hermitian anomalous interactions.  The latter make sense in a generalised PT framework, as has been discussed in \cite{soto} and will be reviewed below. Such non-Hermitian effective actions are characterised by a generalised antilinear $\mathcal C$PT symmetry, with the charge conjugation operation $\mathcal C$ defined appropriately~\cite{antilin,AMS}, to be discussed below, which guarantees the reality of the energy eigenvalues of the system. 

Although the parameters $\alpha^\prime$ and $\kappa^2$ are independent in generic string models~\cite{pol1,pol2}, especially in view of the possibility of large-extra-dimension compactifications, nonetheless for concretreness
in what follows, we shall set~\cite{alex1,soto}
\be\label{alphakappa}
\sqrt{\alpha^\prime }  = M_s^{-1} \sim \kappa = \frac{\sqrt{8\,\pi}}{M_P} \sim (2.4 \times 10^{18})^{-1} \, {\rm GeV}^{-1}.
\ee
where $M_{\rm Pl}  \equiv M_P/\sqrt{8\pi} =  2.4 \times 10^{18}$~GeV is the reduced Planck mass in (3+1)-dimensions.
It goes without saying that this restriction on string parameters does not affect the qualitative conclusions of our analysis on mass generation, and one can straightforwardly extend it to include more general cases in which $\sqrt{\alpha^\prime} \ne \kappa$.

\subsection{Inclusion of fermions \label{sec:fermion}}

\medskip

Upon the inclusion of (Dirac) fermions, the torsion interpretation \eqref{torcon} of the KR three-form $\mathcal H$, implies that the curved-space Dirac terms read~\cite{kaloper,decesare,boss,anomalies}:
\begin{align}\label{fermions}
S_{Dirac} &= \,  \int d^4x \sqrt{-g} \, \Big[ \frac{i}{2} \,\Big(\overline \psi_j \gamma^\mu {\overline {\mathcal D}}(\overline \omega)_\mu \, \psi_j - ( {\overline {\mathcal D}}(\overline \omega)_\mu \, \overline \psi_j  )\, \gamma^\mu \, \psi_j \Big) - m^{(j)}\, \overline \psi_j \, \psi_j \Big], \nonumber \\
& =\,  \int d^{4}x\sqrt{-g}\bar{\psi}_j\Big(\frac{i}{2}\Gamma^{a} \stackrel{\leftrightarrow}{\partial_{a}} - m^{(j)} \Big)\psi_j - \int d^{4}x\sqrt{-g} \, ({\mathcal F}_a + B_a)\, \bar{\psi}_j\gamma^{5}\Gamma^{a}\psi_j  \nonumber \\
& \equiv \; S_{Dirac}^{Free} + \int d^{4}x\sqrt{-g}\, (B_{a}  + {\mathcal F}_a)\,J^{5\, a}~,
\end{align}
where  $\overline \omega_{ab\mu}= \omega_{ab\mu} + K_{ab\mu}$,  $K_{abc} =\frac{1}{2} \, ({\mathcal H}_{cab}  - {\mathcal H}_{abc} - {\mathcal H}_{bca}) = - \frac{1}{2} {\mathcal H}_{abc}$, is the generalised spin-connection with (totally-antisymmetric) torsion $\mathcal H$. As before, Latin indices $a, b, c, \dots$ denote tangent-space indices, raised and lowered with the help of the Minkowski metric $\eta^{ab}$ of the tangent space
(at a point with coordinates $x^\mu$) of a space-time with metric $g_{\mu\nu} (x) = e_\mu^a (x) \, \eta_{ab} \, e_\nu^b (x)$, with  $e^{a}_\mu (x)$  the vielbeins and  $e^\mu_a (x)$ their inverse. $\Gamma^a$ is a tangent-space Dirac matrix, such that the space-time Dirac matrices $\gamma^\mu (x)$ are given by $\gamma^\mu (x) = e^\mu_{a} (x) \, \Gamma^a$,
and we used the standard notation for $\overline \chi\stackrel{\leftrightarrow}{\partial_{a}}\psi = \overline \chi \partial_a \psi - \overline{\partial_a \chi}\,  \psi $. The (gravitational) covariant derivative is given by ${\overline {\mathcal D}}_a  = \partial_a  - \frac{\imath}{4} \, \overline \omega_{bca}\, \sigma^{bc}$, $\sigma^{ab} = \frac{\imath}{2}[\Gamma^a, \Gamma^b]$. 

In \eqref{fermions} we have defined the quantities ${\mathcal F}^d  =   \varepsilon^{abcd} \, e_{b\lambda} \,  \partial_a \, e^\lambda_c$,
$B^d= -\dfrac{1}{4}\,\varepsilon_{abc}^{\;\;\;\;\;\;d}\,{\mathcal H}^{abc}$. In flat or Friedmann-Robertson-Walker space time backgrounds, of interest to us in this review, $\mathcal F^a =0$, and  thus it will not play any r\^ole in our analysis.

The axial fermionic current  is given by : $J^{5 \, \mu} = \bar{\psi}_j\,  \gamma^{\mu} \,\gamma^{5}\psi_j$, and
correspondingly $J^{5 \, a} = \bar{\psi}_j \, \Gamma^{a} \,\gamma^{5}\psi_j$, with the repeated index $j=1, \dots, N_f$ summed over the fermionic matter species $\psi_j$ of the model. The matrix $\gamma^5 = i \, \gamma^0 \, \gamma^1 \dots \gamma^3$ is the standard chirality matrix. 
The term involving the interactions of the $b$-field with the axial current in \eqref{fermions} can be partially integrated to give 
\be\label{bj5}
\int d^4x \, \sqrt{-g}\, \frac{\kappa}{2} \, \sqrt{\frac{3}{2}} \, \partial_{\mu}b \, J^{5\mu}  = - 
\int d^4x \, \sqrt{-g}\, \frac{\kappa}{2} \, \sqrt{\frac{3}{2}} \, b(x) \, J^{5\mu}_{\quad ;\mu}
\ee
where the covariant divergence of the axial current is non-zero in a theory of massless chiral fermions with chiral mixed anomalies, as in \eqref{fermions}:
\be\label{axialdiv} 
J^{5\mu}_{\quad ;\mu} \propto \mathcal G(\omega, \mathbf A) \ne 0
\ee
where $G(\omega, \mathbf A)$ is defined in \eqref{modbianchi2}. The proportionality factors involve the number of chiral fermionic degrees of freedom circulating in the anomalous loop~\cite{alvarez}, which is model dependent.
In some models, e.g. the string-inspired cosmology models of \cite{anomalies,basilakos}, one may encounter a cancellation of the gravitational anomalies, during the radiation and matter epochs of the Universe, but chiral and QCD- anomalies (where the gauge field $\mathbf A$ represents the gluon) survive.

It should be understood in what follows that one may extend the above considerations to include chiral as well as  {\it Majorana} chiral fermions~\cite{alex2,soto}, as is the case of the model of \cite{mp}, which will be our main motivation for the non-Hermitian models discussed in this review.

Adding the fermionic action \eqref{fermions} to the bosonic one \eqref{sea2}, implementing the constraint \eqref{delta}, performing the $\mathcal H$-path integration (in Euclidean formalism), and analytically continuing back to 
Minkowski space time, we obtain the following effective action in the ``\textbf{Scheme I}'' (\eqref{propLC}): 
\begin{align}\label{sea6}
S^{\rm eff \, (I)} =&\; \int d^{4}x\sqrt{-g}\Big[ -\dfrac{1}{2\kappa^{2}}\, R + \frac{1}{2}\, \partial_\mu b \, \partial^\mu b -  \sqrt{\frac{2}{3}}\,
\frac{\kappa}{96} \, \partial_\mu b(x) \, {\mathcal K}^\mu
\Big] \nonumber \\
&+ S_{Dirac}^{Free} + \int d^{4}x\sqrt{-g}\, \Big( {\mathcal F}_\mu + \frac{\kappa}{2} \, \sqrt{\frac{3}{2}} \, \partial_{\mu}b \Big)\, J^{5\mu}    - \dfrac{3\kappa^{2}}{16}\, \int d^{4}x\sqrt{-g}\,J^{5}_{\mu}J^{5\mu}  + \dots \Big] + \dots,
\end{align}
where $\mathcal K^\mu$ has been defined in \eqref{modbianchi2}, and is related to the mixed anomalies. 
The $\dots$ in (\ref{sea6}) indicate gauge field kinetic terms, as well as terms of higher order in derivatives, of no direct relevance to us here. The  four-fermion axial-current-current {\it repulsive} term in \eqref{sea6}, is characteristic of Einstein-Cartan fermionic theories on non-Riemannian spaces with torsion~\cite{hehl,shapiro}, which in our case is provided by the field strength of the KR field, $\mathcal H_{\mu\nu\rho}$ (\eqref{torcon}). The action  \eqref{sea6} is {\it Hermitian}.

On the other hand, in the ``\textbf{Scheme II}'' (\eqref{propLCM}), upon using \eqref{btoib}, one arrives at the following 
{\it non-Hermitian (complex)} effective action:
 \begin{align}\label{sea6II}
S^{\rm eff\, (II)} =&\; \int d^{4}x\sqrt{-g}\Big[ -\dfrac{1}{2\kappa^{2}}\, R + \frac{1}{2}\, \partial_\mu b \, \partial^\mu b -  i\, \sqrt{\frac{2}{3}}\,
\frac{\kappa}{96} \, \partial_\mu b(x) \, {\mathcal K}^\mu
\Big] \nonumber \\
&+ S_{Dirac}^{Free} + \int d^{4}x\sqrt{-g}\, \Big( {\mathcal F}_\mu + i\, \frac{\kappa}{2} \, \sqrt{\frac{3}{2}} \, \partial_{\mu}b \Big)\, J^{5\mu}    - \dfrac{3\kappa^{2}}{16}\, \int d^{4}x\sqrt{-g}\,J^{5}_{\mu}J^{5\mu}  + \dots \Big] + \dots,
\end{align}
which is embeddable in a generalised PT-framework~\cite{soto}, and we shall discuss it below.

The reader should have noticed the invariance of the actions \eqref{sea6} and \eqref{sea6II} under the shift symmetry of the KR axion field 
\be\label{shift}
b(x) \, \rightarrow \, b(x) + {\rm constant},
\ee
which is characteristic of axion physics~\cite{Kim1,kim}. In our string model this is associated~\cite{str2,kaloper,witten} with the U(1) gauge invariance \eqref{Bgauge} of the closed-string actions \eqref{sea2} and \eqref{fermions}. 

\subsection{Non-perturbative effects and breaking of axionic shift symmetry \label{sec:instanton}}

\medskip

In the presence of non-Abelian gauge fields in the anomalous couplings of the axions, e.g. QCD-type gluon fields, 
non-perturbative instantons are responsible for the generation of \emph{shift-symmetry-breaking} axion potentials~\cite{Kim1,kim}.  In particular, for a generic axion field $a(x)$ with anomalous coupling to gluons (or in general non-Abelian gauge fields) of the (generic) form:
\be\label{genericaxion}
S_{\rm axion-QCD} \ni \int d^4x \, \sqrt{-g} \, \frac{1}{f_a} a(x) \mathbf F_{\mu\nu} \, \widetilde{\mathbf F}^{\mu\nu} ~,
\ee
where $f_a$ denotes the pertinent axion coupling (with mass dimension +1), and $\mathbf F_{\mu\nu}$ is the non-Abelian gauge-field strength, non-perturbative instanton effects are responsible for generating an axion potential of the form
 \be\label{axionpot}
 \mathcal U(a) = \Lambda_{\rm inst}^4 \, \Big[1 - {\rm cos}\Big(\frac{a(x)}{f_a}\Big)\Big], 
 \ee
where $\Lambda_{\rm inst}$ is the appropriate scale, e.g. the QCD scale ($\Lambda_{\rm QCD} \sim 218~{\rm MeV}$), if the gauge field pertains to gluons.  The potential \eqref{axionpot}, therefore, breaks the shift symmetry \eqref{shift}, which is now restricted to~\cite{Kim1,kim}
\be\label{shiftbreak}
a(x) \, \rightarrow \, a(x) + 2\pi\, f_a~.
\ee
In our string-inspired model, it is the KR field field that may couple to such instanton effects, through the anomalous terms in \eqref{sea6}. In such a case, the corresponding axion coupling is  given by
\begin{align}\label{fb}
f_b \equiv \sqrt{\frac{8}{3}} \, \frac{\kappa}{\alpha^\prime} =  \sqrt{\frac{8}{3}} \, \Big(\frac{M_s}{M_{\rm Pl}}\Big)^2\, M_{\rm Pl} ~,
\end{align}
where $M_{\rm Pl} =  \kappa^{-1}=  2.4 \times 10^{18}$~GeV is the reduced Planck mass in (3+1)-dimensions, defined in \eqref{alphakappa}. Here we keep the $M_s$ and $M_{\rm Pl}$ different in magnitude, following the phenomenological study in \cite{anomalies,basilakos}. We shall come back to applying \eqref{alphakappa} later. We stress that the generation of a potential for the KR field implies that a torsion interpretation of this field is no longer possible in the massive phase. 

From \eqref{fb}, we thus observe that the range of the axion coupling $f_b$ depends on the allowed range of the string mass scale $M_s$. For the model of, e.g., ref.~\cite{anomalies,basilakos}, the allowed range of the latter scale is
\begin{align}\label{msr}
 \MPl \, \gtrsim  \, M_s \gtrsim 10^{-3} \, M_{\rm Pl}~,
 \end{align}
 where the upper bound of $M_s$ corresponds to \eqref{alphakappa}. This implies 
\begin{align}\label{fbr}
{3.9 \times 10^{12} ~{\rm GeV}\lesssim\, f_b \,\lesssim\,3.9 \times 10^{18} ~{\rm GeV}\,.}
\end{align}
On the other hand, the generic axion coupling constant $f_a$ is constrained to lie in the range~\cite{kim}
\begin{align}\label{far}
 10^9~{\rm GeV} < f_a < 10^{12}~{\rm GeV}~.
\end{align}
We note, though, that astrophysical constraints~\cite{astro1,astro2,astro3,marsh,astro4} may extend the upper bound up to $10^{17}~{\rm GeV}$. We thus observe that, even if the astrophysical constraints are ignored,  there is still a marginal overlap (in order of magnitude) between the minimally allowed region of $f_b$, \eqref{fbr}, and the maximally allowed phenomenological region of the QCD axion coupling constant, \eqref{far}.

From \eqref{axionpot}, for the case of the KR axion $b(x)$, we infer that the instanton-induced KR-axion mass is~\cite{basilakos} 
\begin{align}\label{axionmass}
m_b &= \sqrt{\left. \frac{\partial^2 V_b^{\rm QCD}}{\partial b^2}\right|_{b=0}}  = \frac{\Lambda^2_{\rm QCD}}{f_b} = \sqrt{\frac{3}{8}} \, \Big(\frac{\Lambda_{\rm QCD}}{M_s}\Big)^2\,  \MPl
= \sqrt{\frac{3}{8}} \, \Big(\frac{\Lambda_{\rm QCD}}{\MPl}\Big)^2 \, \Big(\frac{\MPl}{M_s}\Big)^2\,  \MPl ~,
\end{align}
which, in view of \eqref{msr}, \eqref{fbr}, lies in the range
\begin{align}
{ 1.17 \times 10^{-11} ~{\rm  eV} \lesssim m_b \, \lesssim \, 1.17 \times 10^{-5}~{\rm  eV} \,,}
\end{align}
well within the range calculated in lattice QCD approaches (see, e.g. ref.~\cite{latticeqcd} and references therein): $m_a \sim 5.7 \, (\frac{10^{12}~{\rm GeV}}{f_a}) \, \times 10^{-6}$~eV. 

The above considerations pertain to the Hermitian effective action \eqref{sea6} in ``\textbf{Scheme I}''. In this review we shall discuss the phenomenology of
models with non-Hermitian anomalous couplings, \eqref{sea6II},  in ``\textbf{Scheme II}''. Such models  correspond to purely imaginary axion coupling constants for which the aforementioned mass generation through instantons is not clear (for, instance, for purely imaginary $f_a = i f_b, \, f_b \in \mathbb R$, in the potential \eqref{axionpot}, there are no mass terms generated, as the corresponding quadratic terms in the axion field $a(x)$ have the wrong sign). It is the point of this review to discuss alternative ways for mass generation for axions (and fermions) dynamically, which is done in sections \ref{sec:dynamical} and \ref{sec:anom}. We use specific non-Hermitian effective actions which, as we shall discuss, complement \eqref{sea6II} by appropriate {\it anti-Hermitian Yukawa interactions} of axions with the fermions~\cite{mp}. The latter constitute the main source of dynamical-mass generation.

Before doing this, though, it is necessary to motivate the pertinent non-Hermitian models carefully within the microscopic string theory framework. This necessitates  some discussion first on the various types of axions existing in string models, which we now proceed to review.

\subsection{Stringy Model-independent (KR) and Model-dependent axions: kinetic mixing and Yukawa interactions  \label{sec:modelaxion}}

\medskip

 In ref.~\cite{mp} it was postulated that non-perturbative physics might also be responsible for a further breaking 
 of the shift symmetry by means of Yukawa interactions between axions and fermions, specifically right-handed neutrinos, in the context of the above-described Hermitian string-inspired model \eqref{sea6} in the ``\textbf{Scheme I}''.  To this end, one exploits the fact that in string theory~\cite{witten,arvanitaki} there are many axion fields~\cite{witten,arvanitaki}, associated with the Kaluza-Klein (KK) zero modes of appropriate $p$-forms in the spectrum of strings compactified to four space-time dimensions, {\it i.e.} formulated on a target-space-time manifold of the form $M_{1,3} \times \mathcal X_6$. Here, $M_{1,3}$ denotes the uncompactified (3+1)-dimensional space time, and $\mathcal X_6$ the extra-dimensional space, assumed to be a smooth compact manifold. For instance, in heterotic string theory~\cite{str2,pol2}, one has the (Neveu-Schwarz(NS)-type)  two-form  $\mathbf{\mathcal B}$ of the Kalb-Ramond field in ten dimensions, which, upon compactification on an appropriate (say Calabi-Yau~\cite{str2}) 
 six-dimensional compact space, $\mathcal X_6$, can be written as: 
\be\label{KRBfield}
\mathbf{\mathcal B}  = B_{\mu\nu} (x) \, dx^\mu \, dx^\nu + \frac{1}{2\pi} \, b^I(x) \, \omega^I_{ij} (z,\bar z) \, dz^i \, d\bar z^j ~, \quad \mu, \nu=0, \dots 3, \quad i, j =1,2,3  
 \ee
where  $z^i$, $i=1,2,3$ are complex coordinates parametrising the compact manifold. 
The $B_{\mu\nu}(x)$ field yields, upon the dualisation procedure associated with the implementation of the corresponding Bianchi identity for its field strength \eqref{modbianchi2} via a Lagrange multiplier, the KR axion $b(x)$, as discussed above. In the language of string theory this is the so-called {\it model independent} axion, as it is present in all string theories. 
The quantities $\omega^I_{ij}(z, \bar z)$, $I=1, \dots h^{1,1}$ in \eqref{KRBfield}, represent harmonic (1,1) forms that depend only on the coordinates of the complex manifold, and are linked to the aforementioned KK zero modes. One uses the normalisation~\cite{witten}
\be\label{normalisation}
\int_{\mathcal C^J} \omega^I = \delta^{IJ} 
\ee
where $\mathcal C^I$ is a 2-cycle in the compact manifold. In other words, the harmonic forms $\omega^I$ span the integer (1,1) cohomology group of the target space~\cite{eguchi}.

The quantities $b^I(x)$, $I=1, \dots h^{1,1}$ represent dimensionless pseudoscalar fields on the uncompactified space-time, and the factor $\frac{1}{2\pi}$ has been inserted so that the fields $b^I(x)$ have a period $2\pi$, as is conventional for axions\footnote{In this formalism, the corresponding axion fields with mass dimension +1 and canonically-normalised kinetic terms in the effective action are given by $f_{b^J} \, b^J$, where $f_{b^J}$ are the corresponding axion couplings, of mass dimension +1, appearing in the (gauge and gravitational) anomalous, CP-violating, interactions of the axion moduli fields~\cite{witten}.}. These four-dimensional fields correspond to the so-called {\it model-dependent axions}~\cite{witten}. 
The kinetic terms of the two-form yield the four-space-time-dimensional kinetic terms of the $b^I(x)$ fields. Indeed, 
to this end, and taking into account the structure and the space-time dependences of the field components in \eqref{KRBfield}, we need only to consider the following components of the field strength $\mathcal H_{MNP}= \partial_{[M}B_{NP]}$ (with capital Greek letters denoting indices referring to the ten-dimensional space-time of the superstring/brane theory, $M,N,P=0, \dots 9$):
$$ \mathcal H_{\mu ij} = \partial_\mu B_{ij} = (\partial_\mu b^I(x)) \, \omega^I_{ij} (z,\bar z)~, \quad \mu=0, \dots 3, \quad i,j=1,2,3~.$$ 
These components are the only  ones associated with the $b^J$-model-dependent-axion terms on the right-hand side of \eqref{KRBfield}, which are non-vanishing (the terms associated with the first, $B_{\mu\nu}$-dependent terms yield, of course, the purely four-dimensional KR antisymmetric field strength \eqref{KRfs}). 

 The pertinent kinetic terms for the axions $b^I(x)$ in (3+1)-dimensions
stem from $\mathcal H_{MNP} \, \mathcal H^{MNP}$-structures in the ten-dimensional effective action, which, upon compactification down to (3+1)-dimensions, yield terms of the form 
{\small \begin{align}\label{tendim}
S_{\rm 10-dim} &\ni \int \sqrt{-g} \, d^4 x \int_{\mathcal X_6} \partial_\mu B_{ij} \, \partial^\mu B^{ij} = \int \sqrt{-g} \, d^4 x  \, \partial_\mu b^I (x) \partial^\mu b^J (x) \,\int_{\mathcal X_6}  \omega_{ij}^I (z,\bar z) \, \omega^{J\, ij} (z,\bar z)~, 
\nonumber \\ & \equiv \int \sqrt{-g} \, d^4 x  \, \partial_\mu b^I (x) \partial^\mu b^J (x) \, \gamma^{IJ}
\quad \mu=0, \dots 3, \quad I, J=1, \dots h^{1,1},
\end{align}}where, for brevity,  we only indicated the structures, omitting numerical coefficients. The reader should observe the non-trivial kinetic mixing $\gamma^{IJ} \,  \ne \,  \delta^{IJ}$ of the model-dependent stringy axions $b(x)^I$.\footnote{Such mixing has been exploited in \cite{shiu1,shiu2} to discuss a widening of the allowed window of stringy-axion coupling constants, and also study large-field axion-induced inflation.}

It is important to stress that the kinetic terms of the model-depenent stringy axions do {\it not} suffer any ambiguities and are the {\it same} in both ``\textbf{Schemes I and II}'', corresponding to standard axion kinetic terms,  since they do not contain the Levi-Citiva tensor density $\varepsilon_{\mu\nu\rho\sigma}$ \eqref{leviC} (with $ \mu,\nu,\rho,\sigma=0, \dots 3$ denoting indices in the (3+1)-dimensional space-time manifold $M_{1,3}$).

The axion coupling for the fields $b^I(x)$ can be determined by looking~\cite{witten} at the (one-loop) counterterms required for the GS anomaly-cancellation mechanism in string theory~\cite{str2}. As a concrete example, we may consider 
the $E_8 \times E_8$ heterotic string, formulated on $M_{1,3} \times \mathcal X_6$, 
with the Standard Model gauge group $SU(3)_c \times SU(2) \times U_Y(1)$ 
embedded, say, in the first $E_8$ group factor. For brevity and concreteness, we take the uncompactified space-time manifold $M_{1,3}$ to be Minkowski flat.  
As shown in \cite{witten}, then, one obtains the following effective-action anomaly terms for the axion-$b^I(x)$ fields:
{\small \be\label{axionanom}
S_{\rm anom~string~axion} = - \, \Big( \frac{1}{16\, \pi^2} \,\int_{\mathcal X_6} \omega^I (z,\bar z) \, \wedge \, \Big[{\rm Tr_1} \mathbf F \wedge \mathbf F - \frac{1}{2} \, \mathbf R \wedge \mathbf R \Big] \Big) \, \int d^4 x \, b^I(x) \frac{1}{16\pi^2} \, {\rm Tr_1} \mathbf F \wedge \mathbf F
\ee}where, in order to arrive at \eqref{axionanom} starting from the original form of the GS counterterms, the Bianchi identity \eqref{modbianchi2} has been used. The term inside the parentheses on the right-hand side of the above relation expresses mixed anomalies in the compact manifold $\mathcal X_6$, with $\mathbf F$ ($\mathbf R$) the appropriate gauge-field (compact-space-$\mathcal X_6$ curvature) two-form over the compact space, and $\wedge$ the appropriate exterior product among differential forms~\cite{eguchi}, as mentioned previously; the trace ${\rm Tr_1}$ pertains to the first $E_8$ gauge group. 

 We also remark at this point that in generic string or D-brane models, compactified or projected to four space-time dimensions,  {\it model-dependent} axion fields $a^I(x)$ are also obtained as KK zero models of other appropriate $p$-form fields, $C_p$, in the string spectrum on $M_{1,3} \times \mathcal X_6$, for instance, the Ramond-Ramond(RR)-type $p=0,2,4$-forms of type IIB string theory, or the $p=1,3$-forms of type IA~\cite{witten}.  Similarly to the NS 2-form $B$ \eqref{KRBfield}, the corresponding {\it model-dependent}, dimensionless, (3+1)-dimensional axion fields are then given by
 \be\label{pformaxion}
 a^I (x) = \frac{1}{2\pi} \int_{\mathcal C^{(p)}_I} \, C_p~,  \quad  I=1, \dots M,
 \ee
where $\mathcal C^{(p)}_I \ \subset \mathcal X_6$  are appropriate homologically-non-equivalent $p$-cycles in the compact manifold, and we normalised again the axion so as to have period $2\pi$. We stress that qualitatively similar anomaly terms to \eqref{axionanom} occur for the axions $a^I(x)$.

In the model of \cite{mp}, we may assume that there is an anomaly cancellation between gauge and gravitational
anomaly terms in the compact manifold $\mathcal X_6$ in \eqref{axionanom}, which means that such terms {\it vanish}. This is a special case, sufficient for our purposes here. In this case, the stringy model-dependent axions are characterised by their kinetic terms only and have {\it no potential}.  

However, even in such a case, there is an induced coupling between the axions $b^I(x)$ (or $a^I(x)$) and anomaly terms in the (3+1)-dimensional effective action, if there is {\it kinetic mixing} of such model-dependent axions  with the model-independent KR $b$-axion field: 
\be\label{mixing}
S_{\rm kin.~mix.} = \gamma_J \int d^4 x \sqrt{-g} \, \partial_\mu b\, \partial^\mu a_J,  \quad J=1, \dots, M,
\ee 
where the repeated index $J=1, \dots M$, is summed over the stringy axion species. One could also include standard QCD axions in such a summation that may co-exist with stringy axions. Such a mixing has been assumed in the effective field theory of  \cite{mp}. 

The term  \eqref{mixing} could be added to \eqref{sea6}, along with the following Hermitian shift-symmetry-breaking Yukawa interactions~\cite{mp} 
\be\label{yuk}
S_{\rm Yulawa} = i\,\sum_{j=1}^{N_f}  \sum_{J=1}^M\ \lambda_{J\, j} \int d^4 x \sqrt{-g} \, a_J \, \overline \psi_j \, \gamma_5 \, \psi_j ~. 
\ee
where $j$ runs over fermion species $N_f$ in the model.  

Before proceeding in examining the consequences of \eqref{yuk} for 
mass generation, which is the main topic of our work here, 
we should make some important remarks, distinguishing the {\it genuinely quantum}
Yukawa structures \eqref{yuk} from {\it classical} non-derivative Yukawa structures associated with derivative  interactions of axions after the use of fermion equations of motion. The latter characterise any axion model, and are used, for instance, in experimental nuclear-physics searches of axions, as they describe interactions  
of axions to nucleons~\cite{kim}. As such, they are also present 
in the {\it classical} effective action of the (string)model-independent KR axion $b(x)$, due to its torsion nature~\cite{kaloper}.  

Let us explain now how such classical Yukawa interactions arise from derivative, shift-symmetry-respecting interactions. To this end, we first recall that a derivative coupling of a generic axion field $a(x)$ to matter fermions has the form ({\it cf.} \eqref{fermions}):
\be\label{effaf}
S_{a-F}^{\rm eff} \ni \int d^4 x \sqrt{-g}  \, \psi_j \gamma^\mu \, \gamma^5 \psi_j \, \partial_\mu a(x) = 
- \int d^4 x \sqrt{-g}\, a(x) \, \nabla_\mu \, (\psi_j \gamma^\mu \, \gamma^5 \psi_j),
\ee
where in the right-hand side we performed partial integration, assuming as usual that the fields and their derivatives vanish in the boundaries of space-time. 
If the fermions are massless and chiral, then there might be quantum anomalies that could spoil the conservation of the chiral current, $\overline \psi_i \, \gamma^\mu \, \gamma^5 \, \psi_j $, so this term yields non trivial anomalous contributions to the effective action. If, however, the fermions $\psi_j$ are massive, of mass $m^{(j)}$, then the divergence of the chiral current is non-zero already classically, as follows from the equations of motion
\be\label{g5div}
\nabla_\mu \, (\psi_j \gamma^\mu \, \gamma^5 \psi_j) = 2i m^{(j)} \overline \psi_j \, \gamma^5 \, \psi_j ~, \quad j=\dots N_f. 
\ee
In an anomalous quantum theory, of course, the right-hand side of \eqref{g5div} also receives anomalous contributions of the form $\mathcal G(\omega, \mathbf A)$ (defined in \eqref{modbianchi2}), see \eqref{axialdiv}.

Substituting \eqref{g5div} onto the effective action term \eqref{effaf}, one then obtains non-derivative Yukawa coupling terms in the effective action of the form \eqref{yuk}. But this constitutes a classical argument, since the fermion equations of motion have been used. 
Par contrast, the structures \eqref{yuk} represent {\it fully quantum} interactions, for which the fermion equations of motion 
should {\it not} be used. This is why we cannot apply \eqref{yuk} to the model-independent KR axion, because the latter, as we have seen, arises from a dualisation procedure of the gauge-invariant (under the symmetry \eqref{Bgauge}) KR field strength $\mathcal H_{\mu\nu\rho}$,  and, as such, it should appear in the local low-energy effective Lagrangian of string theory through its derivatives.\footnote{The reader might be tempted, following \cite{kaloper,witten}, to redefine the field strength $\mathcal H_{\mu\nu\rho}$ by appropriate counterterms involving fermions, so as to ensure the validity of the modified Bianchi identity
\be\label{mb3}
\varepsilon^{\mu\nu\rho\sigma}\, \mathcal H_{[\nu\rho\sigma\, ;\, \mu]} -  \sqrt{-g}\, {\mathcal G}(\omega, \mathbf{A}) -i\, \lambda \, \sqrt{-g}\, \overline \psi \, \gamma^5 \, \psi =0~, \quad \lambda \in \mathbb R~,
\ee
order-by-order in perturbation theory of the (Hermitian) string-inspired effective field theory described by the actions \eqref{sea2} and \eqref{fermions}. As we have mentioned previously, classically \eqref{g5div} one may obtain the structures $\overline \psi \, \gamma^5 \, \psi $ from the divergence of the axial fermion current, related to the anomalies ${\mathcal G}(\omega, \mathbf{A})$, upon use of the fermion equations of motion for massive fermions. By postulating the constraint \eqref{mb3} one ensures the appearance of such structures at a quantum level as independent operators in the effective action. This would formally ensure the presence of Yukawa interactions $\lambda \, b(x) \, \overline \psi (x) \, \gamma^5 \, \psi (x)$ in the string-inspired field-theory action obtained after the dualisation procedure, by means of which one implements the constraint \eqref{mb3} in the path integral of the effective low-energy field theory via the Lagrange-multiplier field $b(x)$. In such a case, under the 
``\textbf{Scheme II}'', one would obtain effective actions with anti-Hermitian anomaly and Yukawa terms for the model-independent KR axion $b(x)$, embeddable directly in a generalised-PT ($\mathcal C$PT) framework~\cite{qft2,antilin,alex1,alex2,soto}. The problem with this approach, however, is that the constraint \eqref{mb3} would imply a redefinition of the field strength $\mathcal H_{\mu\nu\rho}$, which, apart from the standard Chern-Simons local counterterms \eqref{csterms}, required for the GS anomaly-cancellation mechanism~\cite{str2}, would involve additional {\it non-local} fermiomnic counterterms  of the form (up to numerical coefficients, not relevant for our discussion):
$\propto \, i\, \lambda\, \varepsilon_{\mu\nu\rho\sigma} \, \nabla^\sigma \,\frac{1}{\Box}\, (\overline \psi \, \gamma^5 \, \psi ) $, where $\Box$ denotes the gravitationally-covariant D'Alembertian, $\Box \equiv \nabla_\mu \, \nabla^\mu$, with $\nabla_\mu$ the gravitational covariant derivative. Such non-local terms are not acceptable in the framework of local effective field theories obtained from strings, given that the perturbative-scattering-matrix approach that defines the perturbative-string-theory limit~\cite{str1} would not be valid. Hence we shall not consider such a procedure here, but we felt like mentioning it, in case one is prepared to speculate on the existence of such non-local counterterms in a more general quantum-gravity framework.}

\subsection{Non-perturbative effects and various types of dynamical Majorana neutrino masses \label{sec:nonpert}}

\medskip

In \cite{mp} we therefore considered the addition of the non-derivative interactions \eqref{yuk} in the sector of model-dependent axions, which are not related to the aforementioned dualisation procedure of $\mathcal H_{\mu\nu\rho}$. In this respect, we have  offered an alternative interpretation of the origin of such terms. It has been conjectured in \cite{mp} that  instanton or other non-perturbative effects in string/brane theories can generate such shift-breaking Yukawa couplings, which are therefore characterised by Yukawa couplings $\lambda_{J\, j}$ very small in magnitude ({\it cf.} see \eqref{instl} below). 

It is pointed out at this point that non-perturbatively generated masses for right-handed Majorana neutrinos in string/brane theories have been considered in the literature in the past~\cite{ib1,ib2,cve1,cve2}, 
which are associated with (effective (3+1)-dimensional) field operators of the form 
\be\label{neuop}
M_s \, e^{-U} \, \overline N (x) \, N (x) ~, 
\ee
where $M_s$ is the string mass scale, $N$ is the Majorana field of the neutrino, and $U$ denotes linear combinations of complex string moduli, whose imaginary parts correspond to axion-like particles (which thus are characterised by the standard restricted shift symmetry \eqref{shiftbreak}, corresponding to periodic shifts by $2\pi$ for the dimensionless fields; the Majorana-neutrino-mass operator \eqref{neuop} is also invariant under the standard-model gauge group and a U(1)$_{\rm B-L}$ gauge symmetry, where B(L) denotes Baryon (Lepton) quantum numbers). The string/D-brane instanton zero modes imply right-handed neutrino Majorana masses  suppressed by $\propto \sum_r \mathcal D_r \, e^{-S_r}$, where $\mathcal D_r$ are coefficients depending on flavour couplings of the neutrinos, and $S_r$ is the (real part of the) instanton action, for the type-$r$ stringy instanton (we note that there are many types of instantons in string /brane theories~\cite{ib1,ib2,cve1,cve2}). 

The interaction \eqref{yuk} we are conjecturing in \cite{mp}, and reviewing here, is rather different from \eqref{neuop}, as it depends {\it linearly} on the model-dependent axion, and thus is not characterised by a periodic-shift symmetry, but breaks the shift symmetry completely. It remains to be seen whether microscopic mechanisms in concrete string/brane theories exist that yield \eqref{yuk}. This falls beyond the scope of the current work, where the model with this extra interaction is considered phenomenologically~\cite{mp} as a model beyond the standard model where dynamical right-handed neutrino masses can be generated in non-conventional ways.
Moreover, the kinetic mixing \eqref{mixing} itself, 
could be the result of the same non-perturbative physics~\cite{alex1,alex2}, {\it e.g.} through one loop-effects involving heavy fermions, existing only in internal lines of graphs, coupled to axions $b$ and $a^I$ via the (non-perturbatively generated) Yukawa couplings \eqref{yuk}.
 
 In the model of \cite{mp}, the non-perturbative Yukawa coupling couples to right-handed Majorana neutrinos $\psi_R$ only, hence one considers adding the following Hermitian terms in  the Hermitian effective string action \eqref{sea6}, within the ``\textbf{Scheme I}'' framework:
{\small \begin{eqnarray} \label{bacoupl3} 
\mathcal{S}_{a-b} = \int d^4 x
    \sqrt{-g} \, \Big[\frac{1}{2} (\partial_\mu a(x) )^2  + \gamma  \, \partial_\mu b\, \partial^\mu a     +  i\, \lambda \, 
a(x)\, \Big( \overline{\psi}_R^{\ C} \psi_R - \overline{\psi}_R
\psi_R^{\ C}\Big)  
    + \dots \Big]\;, \nonumber \\
\end{eqnarray}}where $\psi_R^{\ C} = (\psi_R)^C$ is the (Dirac) charge-conjugate of the right-handed fermion $\psi_R$,
which in \cite{mp} is considered to be a sterile neutrino $\nu_R$ (the reader should notice that, due to the chiral nature of the fermions $\psi_R$ , the $\lambda$-dependent terms in \eqref{bacoupl3} correspond to the axial form in \eqref{yuk}). In \eqref{bacoupl3}, for brevity, we did not write explicitly the kinetic terms of $\psi_R$. These are understood to be included in the $\dots$. Due to the non-perturbative nature of $\lambda$ one may assume 
\be\label{instl}
\lambda \sim \exp(-S_{{\rm instanton}})\, \ll\, 1, 
\ee
where $S_{{\rm instanton}} \gg 1$ is the (real part of an appropriate) large and positive instanton action.\footnote{In view of the above discussion on the existence of many types of instantons in string theory, one might consider $\lambda$ as proportional to the appropriate sum involving many exponential suppression factors of the (real parts of the) actions of individual instantons.}

Redefining the $b$-field as $b \rightarrow b^\prime = b + \gamma \, a $, one diagonalises the kinetic axion terms in \eqref{bacoupl3}, and can  readily perform the 
$b^\prime $ field integration in the path integral corresponding to the effective action obtained from the sum of \eqref{sea6} and \eqref{bacoupl3}. In this way one arrives at an effective (3+1)-dimensional action for the dynamics of the stringy axion field $a(x)$~\cite{mp}. Upon rescaling appropriately the $a(x)$ field, so as to have canonical kinetic terms after the above procedure, one arrives at the following effective action involving the $a$-field:
\begin{eqnarray} \label{aeff} 
\mathcal{S}_{a}^{\rm eff} \Big|_{\rm Scheme~I} \!&=&\! \int d^4 x
    \sqrt{-g} \, \Big[\frac{1}{2} (\partial_\mu a(x) )^2  - \frac{\gamma\, 
        a(x)}{f_b\, \sqrt{1 - \gamma^2}}\, \mathcal G (\omega, \mathbf A)
       \nonumber\\ 
&&\hspace{-5mm} + \frac{i\, \lambda}{\sqrt{1 - \gamma^2}} \, 
a(x)\, \Big( \overline{\psi}_R^{\ C} \psi_R - \overline{\psi}_R
\psi_R^{\ C}\Big)  
    +  \dots \Big]\;,
\end{eqnarray}
where the KR-axion coupling $f_b$ is defined in \eqref{fb}. 
The $\dots$ include terms independent of $a(x)$, including the kinetic terms of the $\psi_R$ fermions, 
as well as anomaly-dependent terms stemming from the 
$b^\prime (x)$-field path integration, of the form $\mathcal K_\mu \mathcal K^\mu$, 
with $\mathcal K^\mu$ defined in \eqref{modbianchi2}.  The reader should notice the induced 
anomalous interactions of the (model-dependent stringy) axion field $a(x)$ field in \eqref{aeff}, proportional to the kinetic-mixing parameter $\gamma/\sqrt{1-\gamma^2}$.
It should be stressed that, to avoid the axion behaving as a ghost, one needs to impose the condition~\cite{mp}
\be\label{condgamma}
0 < \gamma^2 < 1 ~.
\ee
\begin{figure}[t]
 \centering
  \includegraphics[clip,width=0.40\textwidth,height=0.15\textheight]{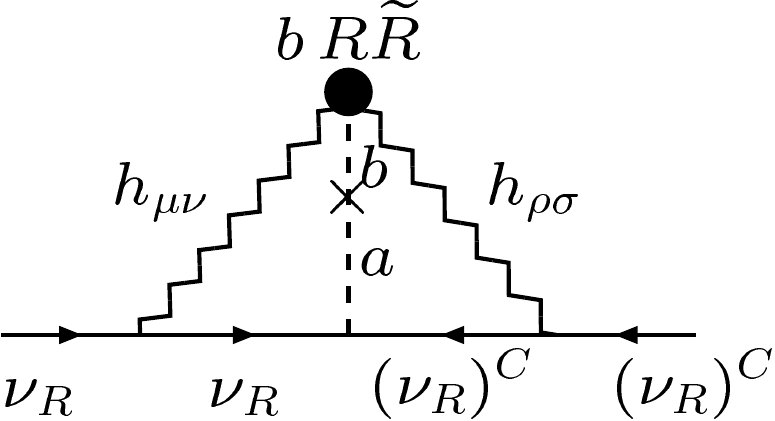} 
\caption{\it Feynman graph pertaining to the anomalously generated 
 Majorana mass for the right-handed fermions $\psi_R$, identified with sterile neutrinos $\nu_R$ in \cite{mp}.  The dark blob denotes the operator $b(x)\, R_{\mu\nu\lambda\rho}\widetilde{R}^{\mu\nu\lambda\rho}$ (up to numerical coefficients), which is the only part of the anomaly relevant for sterile neutrinos, which do not couple to gauge fields; 
 $b(x)$ denotes the KR (gravitational) axion and $a(x)$ an axion-like particle. The wavy lines are gravitons $h_{\mu\nu}$, and dashed lines denote axions. The cross ``$\times$'' indicates the b(x)-a(x) kinetic mixing \eqref{mixing}. Figure taken from \cite{mp}.}
\label{fig:anom}
\end{figure}
From the effective action \eqref{aeff}, one may deduce novel ways of generating {\it radiatively} Majorana masses for the right-hended fermions (sterile neutrinos in the model of \cite{mp}), which are based on diagrams depicted in figure \ref{fig:anom}. The radiatively-generated Majorana sterile-neutrino mass is $M_R$ found to be~\cite{mp}:
\begin{align}\label{massR}
M_R \sim \frac{\sqrt{3}\, \lambda\, \gamma\, \kappa^5 \Lambda^6}{49152\sqrt{8}\,
\pi^4 (1 - \gamma^2 )}\; ,  
\end{align}
where $\Lambda$ is the Ultra-Violet (UV) momentum cutoff. The computation involved the use of graviton fluctuations $h_{\mu\nu}(x)$ about flat Minkowski space-time backgrounds. These yield non-trivial contributions to the 
gravitational-anomaly term $R_{\mu\nu\rho\sigma} \, {\widetilde R}^{\mu\nu\rho\sigma} $, which are computed using perturbative (quantum) gravity methods developed  in \cite{donogh} (note that the gravitational-anomaly term vanishes for flat or cosmological Friedmann-Lemaitre-Robertson-Walker background space-times).
In the context of string theory, 
$\Lambda$  and  $\kappa^{-1}$  are related~\cite{mp} via the string mass scale and compactification radii of the extra-dimensional spatial manifold. 
For a generic quantum gravity model, independent of string theory, one may use simply  $\Lambda \sim \kappa^{-1}$. The sterile-fermion mass \eqref{massR} is {\it independent} of the axion-$a(x)$ potential~\cite{mp}, and thus its mass. As the reader can notice, the mass $M_R$ is generated for arbitrarily small $\gamma$ and $\lambda$ (real) couplings. This covers the case where both $\gamma$ and $\lambda$ are generated non-perturbatively, say by instanton effects, as discussed previously, and thus are very small, of order \eqref{instl}.

We now remark that, in the ``\textbf{Scheme II}'', in which there are purely-imaginary anomalous couplings of the 
KR model-independent axions $b(x)$ in the effective action \eqref{sea6II}, a redefinition of the KR axion field
$b \rightarrow b^\prime = b +  i \, \gamma \, a $, decouples the model-independent axion from the model-dependent one, $a(x)$, and, as the reader can easily verify, upon integrating out formally in the path integral $b^\prime(x)$, one obtains the following effective action for the model-dependent axion field $a(x)$:
\begin{eqnarray} \label{aeff2} 
\mathcal{S}_{a}^{\rm eff} \Big|_{\rm Scheme~II} \!&=&\! \int d^4 x
    \sqrt{-g} \, \Big[\frac{1}{2} (\partial_\mu a(x) )^2  + \frac{\gamma\, 
        a(x)}{f_b\, \sqrt{1 + \gamma^2}}\, \mathcal G (\omega, \mathbf A)
       \nonumber\\ 
&&\hspace{-5mm} + \frac{i\, \lambda}{\sqrt{1 + \gamma^2}} \, 
a(x)\, \Big( \overline{\psi}_R^{\ C} \psi_R - \overline{\psi}_R
\psi_R^{\ C}\Big)  
    +  \dots \Big]\;,
\end{eqnarray}
where the $\dots$ include anomaly-dependent terms stemming from the $b^\prime$ integration of the form $-\mathcal K^\mu \mathcal K^\mu$ (note the opposite sign as compared with the corresponding terms in the Hermitian  case \eqref{aeff}). The reader should notice that there is no restriction in the range of $\gamma$ in this case. It should also be noticed that the action \eqref{aeff2} is still Hermitian, despite the non-Hermiticity of the KR-$b$-axion Lagrangian \eqref{sea6II}, in ``\textbf{Scheme II}''. Through the anomalous (radiative) graphs of fig.~\ref{fig:anom}, one generates a right-handed-neutrino Majorana mass in this case of the form
\be\label{massR2}
M_R \sim \frac{\sqrt{3}\, \lambda\, \gamma\, \kappa^5 \Lambda^6}{49152\sqrt{8}\,
\pi^4 (1 + \gamma^2 )},
\ee
for the entire range of $\gamma \in \mathbb R$ (in computing the graph of fig.~\ref{fig:anom} care should be taken so that the product $\frac{\lambda \, \gamma }{\sqrt{1+\gamma^2}}$ is not too big, otherwise our perturbative analysis~\cite{mp} breaks down; this is guaranteed for sufficiently small $|\lambda \, \gamma| < 1$). 

A final, but important, comment, we would like to make before closing this subsection, concerns the motivation for the absence  of any potential term for the axions in our studies here and in \cite{alex1,alex2,soto} ({\it cf.} \eqref{aeff}  and \eqref{aeff2}).   As remarked above, the perturbative (radiative) mass generation for the masses \eqref{massR} (or \eqref{massR2}) does not depend on the details of the axion-$a$ potential. This is to be contrasted~\cite{alex1,alex2}, though, with the dynamical SD mass due to the Yukawa couplings \eqref{yuk}  which depends on the self-interactions of the axion field (see discussion in section~\ref{sec:nh4f}). Such potentials, as we have already discussed in section \ref{sec:modelaxion}, can be generated by instanton effects at some energy scale. In our previous discussion, however ({\it cf.} text following \eqref{pformaxion}), we have seen that, by {\it e.g.} fine-tuning anomalies  in the compact sector of string theories, (\eqref{axionanom}), one could cancel such potential terms for model-dependent axions. One can extend the absence of instanton-induced potentials for the KR model-independent axion as well, as happens in some string-inspired cosmological models~\cite{anomalies,basilakos}. There, one may assume that the energy scale of the instanton effects is such that the generation of the pertinent potential takes place at, say, the radiation epoch of the string Universe, and thus the KR axion is massless, without self interactions, during the entire inflationary period. In such a context, our {\it prototype} models of axions without potentials, which are also embeddable in a PT-symmetry framework, to be discussed below, could play a r\^ole as providers of alternative scenarios for dynamical-mass generation for axions and fermions at such early stages of the string-inspired Universe~\cite{alex2}.

\subsection{Motivation for embedding  the model \eqref{aeff2} in a generalised-PT ($\mathcal C$PT) framework \label{sec:pt}}

\medskip

We next remark that, although from a string point of view, the resulting effective actions involving model-dependent axions and non-derivative Yukawa interaction with fermions are real ({\it cf.} \eqref{aeff}, \eqref{aeff2}), nonetheless we {\it may embed} them in a non-Hermitian PT-symmetry framework by {\it analytically continuing both} coefficients $\gamma$ and $\lambda$ to purely imaginary values~\cite{alex1,alex2}
\be\label{analytically}
\gamma \, \rightarrow i\, \gamma, \quad \lambda \, \rightarrow \, i\, \lambda, \quad \gamma,\, \lambda \, \in \, \mathbb R.
\ee 
Formally, such a procedure will leave the radiatively generated masses \eqref{massR}, \eqref{massR2} {\it real}, but now the corresponding effective actions will have non-Hermitian interactions. This result, however, is only formal, given that the presence of {\it non-Hermitian anomaly terms} are yet to be understood. Hence, the methods developed for the Hermitian case to deal with graviton fluctuations of the gravitational anomaly terms~\cite{donogh} which lead~\cite{mp} to the radiative mass \eqref{massR} might not apply. 

We therefore seek for alternative mass generation in such non-Hermitian cases embeddable in a generalised PT-symmetric framework~\cite{qft2}, which is actually~\cite{alex1,alex2,soto} $\mathcal C$PT symmetric, under an appropriate definition of the charge-conjugation $\mathcal C$ operator~\cite{AMS}, as we shall discuss in the next section. As already mentioned, such an embedding guarantees the reality of the pertinent energy spectra~\cite{antilin} and, thus, potentially of the masses that can be generated dynamically. The motivation for the analytic continuation \eqref{analytically}, lies on the fact that the Hermitian Yukawa interactions
cannot lead to dynamical generation of axion and fermion masses, unless there is a {\it bare} axion mass present~\cite{alex2,soto}. Par contrast, in the non-Hermitian case \eqref{analytically} one can obtain {\it truly dynamical} masses for {\it both} axions and fermions, in the absence of bare masses for such fields, but in the presence of attractive four-fermion interactions, as we shall discuss below. The physical interest in having massive axions is linked with the potential r\^ole of such fields as dark matter.

It is the purpose of this article to review briefly the dynamical generation of masses for axions and fermions in this latter (non-Hermitian) set up. In the next section \ref{sec:dynamical}, we examine the issue of dynamical mass generation in theories with non-Hermitian (actually, anti-Hermitian) Yukawa couplings in the absence of anomalous terms, i.e. $\gamma=0$, $\lambda \ne 0$. Motivated by our arguments about the potential generation of such Yukawa interactions through non-perturbative effects, we restrict ourselves to very small Yukawa coupling $|\lambda| \ll 1$. We incorporate a non zero $\gamma \ne 0$, $|\gamma < 1|$, as well as non-Hermitian axial backgrounds, in section \ref{sec:anom}. Despite the smallness of the couplings $\lambda$ and $\gamma$, we use non-perturbative Schwinger-Dyson (SD) analysis.

In what follows we shall use only {\it prototype models}, involving one Dirac fermion and one axion, without potential for the axion field, accroding to our previous discussion. This suffices to demonstrate the main conclusions. 
The extension to Majorana fermions is straightforward and has been done in \cite{alex2,soto}, where we refer the 
interested reader for details. These references also contain an analysis of SD dynamical mass generation for the Hermitian effective actions \eqref{aeff} (or \eqref{aeff2}). 
 
We remark at this point that, in addition to the $\lambda$- and $\gamma$-dependent  terms in \eqref{aeff} (or \eqref{aeff2}), which are 
analytically continued, as per \eqref{analytically}, to yield the corresponding anti-Hermitian, $\mathcal C$PT-invariant theories of interest to us here~\cite{alex1,alex2,soto}, we shall also include {\it attractive Hermitian} four-fermions interactions. Such interactions are present in any low-energy effective string theory with fermions, as a consequence of the exchange of heavy string states among fermions~\cite{benakli}.  Integrating out such heavy states results in effective low-energy actions with contact four-fermion interactions of various types. It suffices for our purposes to consider the addition to our prototype models of four-fermion attractive interactions of the form 
 \begin{align}\label{4feff}
 -\frac{1}{2\,f_4^2} \Big(\overline {\psi}\, \gamma^5 \, \psi\Big)^2~,
\end{align}
where $f_4$ is an appropriate coupling with dimensions of [mass]. As we shall discuss below, the r\^ole of such four-fermion interactions is crucial for fermion dynamical mass generation. Therefore, in view of the 
quantum-torsion-induced four-fermion {\it repulsive} interactions involving the covariant square of the axial fermion current $J_\mu^5$ in \eqref{sea6} (or \eqref{sea6II}), which upon use of Fierz identities, yield - among other terms -  repulsive structures of the form \eqref{4feff}, one must have {\it sufficiently strong} attractive interactions \eqref{4feff} to {\it overcome} the corresponding torsion-induced {\it repulsions}. This will be assumed in what follows, and from now on we ignore the repulsive four-fermion interactions due to torsion, as we assume that the dimensionful coupling $f^{-2}_4$ of the interaction \eqref{4feff} describes an effective interaction where the effects of repulsion have been subtracted appropriately. 

As we shall see in the following sections, the dynamically-generated masses will be much smaller than the ultraviolet (UV)  momentum cut-off scale $\Lambda$ in the effective, string-inspired, low-energy four-dimensional theory~\cite{mp,alex1,alex2}. It is natural to consider $\Lambda \lesssim M_s$, where $M_s$ is the string scale, above which stringy effects appear, which would jeopardize the validity of any point-like string-inspired effective action. As we have demonstrated in \cite{alex2,soto}, and shall review below, the four-fermion dimensionful coupling $f_4$ and UV cut-off scale $\Lambda$, will be determined self-consistently by the SD analysis in the massive phase of the models~\cite{alex2}.

\medskip 

\medskip

\section{Dynamical Mass generation induced by non-Hermitian Yukawa interactions \label{sec:dynamical}}

\medskip

We consider the following model consisting of a single pseudoscalar (axion) field $\phi(x)$ and a single Dirac fermion,
$\psi (x)$, interacting with each other via a non-Hermitian Yukawa (specifically, anti-Hermitian)  interaction with a real coupling $\lambda \in \mathbb R$. Both axion and fermion fields have zero bare masses.  We restrict ourselves to Minkowski space time
for concreteness. The action reads
\be\label{actionnhyuk}
\mathcal S_{\rm Yuk}^{\rm non-Herm} = \int d^4 x \Big(\frac{1}{2} \,\partial_\mu \phi \, \partial^\mu \phi + \overline \psi \, i \, \slashed{\partial} \, \psi + \lambda \, \phi \, \overline \psi \, \gamma^5 \, \psi  - \frac{g^2}{2\, f_4^2} \,  (\overline \psi \gamma^5 \, \psi )^2 \Big), \quad \lambda \in \mathbb R, 
\ee
where we use the standard notation $\slashed{\partial} \equiv \gamma^\mu \, \partial_\mu$, with $\gamma^\mu$, $\mu=0, \dots 3$, the four-dimensional Dirac matrices. Following the discussion after \eqref{pformaxion} in section \ref{sec:modelaxion}, we include {\it no axion-$\phi$ potential} in the model. 
The model is $\mathcal C$PT invariant, provided one uses the following definitions for the discrete transformations for
parity P, time-reversal T and charge conjugation $\mathcal C$, acting on fermions~\cite{alex1}:
\bea\label{cpt}
\mbox{parity}~~~~P&:&~~~~\psi(t,\vec r)\to\gamma^0\psi(t,-\vec r)\nonumber \\
\mbox{time-reversal}~~~~T&:&~~~~\psi(t,\vec r)\to \gamma^1\gamma^3\psi(-t,\vec r) \nonumber \\
\mbox{charge conjugation}~~~~ \mathcal C&:&~~~~\psi(t,\vec r)\to-i\gamma^2\psi^\ast(t,\vec r)~,
\eea
such that the anti-Hermitian interaction $\phi\,\overline \psi\, \gamma^5\, \psi$ is ${\mathcal C} PT$-even under the above transformations if $\phi$ is a pseudo-scalar.\footnote{We note for completion that, in case $\phi$ is a scalar, one can also guarantee the $\mathcal C$PT invariance of the pertinent action, by defining appropriately the corresponding $\mathcal C$ operator.}  It is the existence of this anti-linear ${\mathcal C}PT$ symmetry that guarantees the existence of real energy eigenvalues, and thus real dynamical masses~\cite{alex1,alex2}, for this anti-Hermitian system, according to the general arguments of \cite{most2,most3,most,maninnerPT,antilin}. 

In \eqref{actionnhyuk} the parameter $g$ is set to take on the values $g=0,1$, depending on whether we consider the absence or the presence of attractive four-fermion interactions, respectively, and $f_4$ is an effective  dimensionful coupling, with mass dimensions of [mass], which is determined self-consistently in a SD analysis~\cite{alex2}, as we shall review below. As already mentioned, in the context of microscopic string-inspired models,  the attractive four-fermion interactions in \eqref{actionnhyuk} express the net result, after appropriate subtraction of the corresponding repulsive interactions of the same form, obtained after applying the Fierz identities to the repulsive axial-current-current four-fermion interaction $\frac{3\kappa^2}{16}\, J_\mu^5 \, j^{5\,\mu}$ in \eqref{sea6II}. Indeed, for  a single fermion we are considering here, one has\footnote{The reader should have noticed that, upon bringing the vector-current-current interaction onto the left-hand-side of \eqref{fierz}, one would obtain a combination of four-fermion interactions which, if attractive, would coincide with the chiral-symmetry-invariant Nambu-Jona-Lasinio four-fermion interactions~\cite{NJL}. However, as we have seen above, the quantum torsion generates only the axial-current-current part of this and is repulsive. In string effective theories~\cite{benakli}, as already mentioned, one has various vector-current-current induced terms with coefficients inversely proportional to the square of the mass of the exchanged massive string states. In our work here we are not interested in a detailed study of such four-fermion interactions. In this review, we shall only use \eqref{fierz}  to make some generic phenomenological observations on the magnitude of $f_4$, consistent with SD mass generation, which would allow us to ignore the repulsive torsion-induced interactions.}
{\small \be\label{fierz}
\frac{3\kappa^2}{16}\, J_\mu^5 \, j^{5\,\mu} =  \frac{3\kappa^2}{16}\, (\overline \psi \, \gamma^5 \, \gamma^\mu \, \psi )\, (\overline \psi \, \gamma^5 \, \gamma_\mu \, \psi ) = \frac{3\kappa^2}{16}\,
(\overline \psi  \, \gamma^\mu \, \psi )\, (\overline \psi  \, \gamma_\mu \, \psi )  +  \frac{3\kappa^2}{8}\, (\overline \psi \, \psi )^2 - \frac{3\kappa^2}{8}\ \,(\overline \psi \, \gamma^5\, \psi )^2.
\ee}
We shall come back to this point later on in the article, after we estimate the strength of $f_4$ necessary for a consistent  SD dynamical mass generation~\cite{alex2}.

At the moment we remark that, motivated by the stringy axion models discussed above, we consider very small Yukawa couplings $|\lambda | \ll 1$ in \eqref{actionnhyuk}.
We are interested in dynamical mass generation, due to the coupling $\lambda$, as a way of mass generation alternative to the radiative anomalous generation \eqref{massR2}, and independent of the anomalous coupling $\gamma$, since in the model \eqref{actionnhyuk} $\gamma=0$. We shall follow a Schwinger-Dyson (SD) approach to study dynamical mass generation for both axions and fermion fields~\cite{alex1}. 

Before doing this, we would like to remark that in the absence of attractive four-fermion interactions, i.e. when $g=0$, there is {\it no} dynamical mass generation for {\it fermions}, but there could be for axions (pseudoscalar) $\phi(x)$ fields, as follows from {\it energetics arguments}~\cite{alex1} that we review in the next subsection.

\subsection{Energetics arguments against dynamical generation of fermion mass in non-Hermitian Yukawa models without four-fermion self interactions \label{sec:ener}}

\medskip 

To this end~\cite{alex1}, let us first assume a non-zero {\it bare} mass for the scalar $M_0 \ne 0$. This will serve as a regulator for the definition of inverse propagators, as discussed in \cite{alex1}. Our arguments on mass generation will not be affected by its presence, and thus $M_0$ can eventually be removed, as we shall see. To discuss dynamical mass generation for the fermions in non-Hermitian models a Euclidean path-integral quantisation 
is necessary (with a Euclidean-metric signature convention (+,+,+,+), which is used throughout this work), in which the 
anti-Hermitian Yukawa interaction appears as a {\it phase}:
\be
Z_\lambda[j,\bar\eta,\eta]=\int{\cal D}[\phi,\psi,\bar\psi]\exp\left(-S_{Herm}-S_{anti-Herm}-S_{sources}\right)~,
\ee
where
\bea\label{defS}
S_{Herm}&=&\int d^4x\left(\frac{1}{2}\partial_\mu\phi\partial^\mu\phi+\frac{M_0^2}{2}\phi^2+\bar\psi i\slashed\partial\psi \right), \quad
S_{sources} = \int d^4x (J\phi+\bar\eta\psi+\bar\psi\eta),\nonumber \\
S_{anti-Herm}&=&-\lambda\int d^4x~\phi\bar\psi\gamma^5\psi~=~i\lambda\int d^4x ~\phi \, \Phi
~~~~\mbox{with}~~~~\Phi\equiv \mbox{sign}(i\bar\psi\gamma^5\psi)|\bar\psi\gamma^5\psi|~.\nonumber
\eea
The source-$J,\eta, \bar \eta$ terms in \eqref{defS} have been inserted for the correct definition of the path integral. 

From a basic property of complex calculus we obtain the inequality~\cite{alex1}
\bea\label{vwar}
\exp(-S^{ferm}_{eff})&\le&\int{\cal D}[\phi]~
\Big|\exp\left(-\int_x~\bar\psi i\slashed\partial\psi +\bar\eta\psi+\bar\psi\eta+\frac{1}{2}\phi G^{-1}\phi
+i\lambda\phi\Phi\right)\Big| \nonumber \\
&=&\int{\cal D}[\phi]~\exp\left(-\int_x~\bar\psi i\slashed\partial\psi +\bar\eta\psi+\bar\psi\eta
+\frac{1}{2}\phi G^{-1}\phi\right)~,
\eea
We note that $S^{ferm}_{eff}$ plays the role of the (Euclidean) {\it vacuum energy functional}. hence, as a 
consequence of \eqref{vwar}, it appears that this quantity 
is {\it larger} in the non-Hermitian Yukawa theory than in the free theory. This precludes energetically 
fermion dynamical mass generation~\cite{VW},  
in contrast to the usual Hermitian case, where such a dynamical mass lowers the energy of the system~\cite{alex2}.
This can also be seen explicitly by integrating out the massive scalar field $\phi$. In such a case, one obtains an effective fermionic action
$S^{ferm}_{eff}$:
\bea
\exp(-S^{ferm}_{eff})&\equiv&\exp\left(-\int_x~\bar\psi i\slashed\partial\psi+m\bar\psi\psi+\bar\eta\psi+\bar\psi\eta\right)
\int{\cal D}[\phi]\exp\left(-\int_x\frac{1}{2}\phi G^{-1}\phi+i\lambda\phi~ \Phi\right) \nonumber \\
&=&\exp\left(-\int_x~\bar\psi i\slashed\partial\psi+m\bar\psi\psi+\bar\eta\psi+\bar\psi\eta
+\frac{\lambda^2}{2}\Phi G\Phi\right)~,
\eea
where $G^{-1}=-\Box+M_0^2$ and $\Phi$ is defined in (\ref{defS}) (here the reader recognises the importance of assuming a non-zero bare mass for the scalar, $M_0 \ne 0$, so as to have a well-defined operator $G^{-1}$). Ignoring higher-order derivatives, 
which are irrelevant for our analysis, we obtain
\be\label{Sefffermion}
S^{ferm}_{eff}\simeq\int_x~\bar\psi i\slashed\partial\psi+m\bar\psi\psi+\frac{\lambda^2}{2M_0^2}|\bar\psi\gamma^5\psi|^2~.
\ee
The reader should then notice that this effective action includes a {\it repulsive} four-fermion interaction, for all values of the regulator $M_0$, which increases the energy of the system, in agreement with the 
generic argument \eqref{vwar}, and thus
dynamical mass generation for the fermions is impossible.  Moreover, the form of the induced repulsive interaction implies the absence also of the chiral mass $\mu$, as the latter would be associated with a condensate of the form
$<i\overline \psi \, \gamma^5 \, \psi >$, which is also not formed due to the above energetics reasons.

Par contrast, one may have axion mass generation, which can be established upon considering 
the effective action $S^{scal}_{eff}$ for the pseudoscalar field $\phi$, obtained after integrating out massive fermions (of bare mass $m_0 \ne 0$)
\be
\exp(-S^{scal}_{eff})\equiv\exp\left(-\int_x\frac{1}{2}\partial_\mu\phi\partial^\mu\phi+\frac{M_0^2}{2}\phi^2+J\phi\right)
\int{\cal D}[\psi,\bar\psi]\exp\left(-\int_x\bar\psi(i\slashed\partial+m_0-\lambda\phi\gamma^5)\psi\right)~.\nonumber
\ee
For a constant scalar field configuration $\phi_0$, which suffices for demonstrating our main argument, the effective potential reads~\cite{alex1}
\be
U_{eff}(\phi_0)=\frac{M_0^2}{2}\phi_0^2- \mbox{Tr}\left\{\ln(\slashed p+m_0-\lambda\phi_0\gamma^5)\right\}~,
\ee
which implies
\be\label{dUeff}
\frac{dU_{eff}}{d\phi_0}
=M_0^2\, \phi_0 + \frac{\lambda^2\phi_0}{4\pi^2}
\left(\Lambda^2-(m_0^2-\lambda^2\phi_0^2)\ln\left(\frac{\Lambda^2+m_0^2-\lambda^2\phi_0^2}{m_0^2-\lambda^2\phi_0^2}\right)\right)~,
\ee
where $\Lambda$ is the UV cut off. 
The energies are therefore {\it real}, and the non-Hermitian theory is self consistent, for $m^2\ge\lambda^2\phi_0^2$, as expected from the study in \cite{qft2}. The possibility of real energies is attributed, of course, to the underlying $\mathcal C$PT antilinear symmetry~\cite{antilin,alex1}. 
In the limit $\lambda^2\phi_0^2\to m^2$, the effective potential 
becomes a mass term 
\be\label{Ueff}
U_{eff}\to \frac{1}{2}(M^{(1)})^2\phi_0^2~~~~\mbox{with}~~~~(M^{(1)})^2=M_0^2+\frac{\lambda^2}{4\pi^2}\Lambda^2~.
\ee
From eq.(\ref{Ueff}) it is evident that dynamical axion-mass generation is
possible in the anti-Hermitian-Yukawa-interaction model \eqref{actionnhyuk} (with $g=0$), since on setting the bare mass to zero, $M_0 =0$, one obtains  a non-zero result from \eqref{Ueff} 
\begin{align}\label{dynam}
(M^{(1)})^2_0 = \frac{\lambda^2}{4\pi^2}\Lambda^2 > 0,
\end{align}
which is the same as the result of a one-loop calculation. 

In \cite{alex2} the same conclusions as above  are reached by a detailed SD analysis, which avoids the use of 
regulator bare masses. We refer the interested reader to that work for details. Below we shall only describe the main results of such an analysis.

\subsection{Schwinger-Dyson Analysis of non-Hermitian Yukawa interactions, in the absence of four-fermion interactions \label{nhyukno4f}}

\medskip

To formulate properly our quantum non-Hermitian prototype models, one necessarily uses, as already mentioned,  a Euclidean (``$E$'') path-integral formalism. For the model \eqref{actionnhyuk} with $g=0$, of interest in this subsection, the appropriate Euclidean generating functional is given by 
\begin{align}\label{partf}
  Z^E [J, \eta, \overline{\eta}] = \int \mathcal{D} [\phi  \psi
  \overline{\psi}] \exp \left\{ - \int d^4 x  \left[ \frac{1}{2} \, \partial_{\mu} \phi
  \partial^{\mu} \phi  + \overline{\psi} i\slashed{\partial} \psi - \lambda
  \phi \,  \overline{\psi} \gamma^5 \psi \right] - \int d^4 x [J
  \phi  + \overline{\psi} \eta + \overline{\eta} \psi] \right\}.
\end{align}

Taking into account that we are dealing with very small Yukawa couplings, $|\lambda| \ll 1$, possibly generated 
by non-perturbative effects in the underlying microscopic string/brane theories, and following standard SD methods,
one arrives at the  following SD equations for the pseudoscalar ($s$) and fermion ($f$) propagators in Fourier space~\cite{alex2}:
\begin{align}\label{SD1}
  G^{- 1}_s (k) - S^{- 1}_s (k) &=  \lambda \tmop{tr} \left[ \gamma^5 \int \frac{d^4 p}{(2 \pi)^4} G_f (p) \Gamma^{(3)} (p, k) G_f (p - k) \right] \nonumber \\
  G^{- 1}_f (k) - S^{- 1}_f (k) &= - \lambda \gamma^5 \left( \int \frac{d^4 p}{(2 \pi)^4} G_f (p) \Gamma^{(3)} (p, k) G_s (p - k) \right),
\end{align}
where the dressed inverse propagators are given by: $G_f^{-1}(p)=\slashed{p}+m+\mu \gamma^5$, $G_s^{-1}(p)=p^2+M^2$ and we used the bare inverse propagators $S_f^{-1}(p)=\slashed{p}$, $S_s^{-1}(p)=p^2$. 
The vertex function is in the rainbow approximation~\cite{alex2}, appropriate for weak Yukawa interactions, $|\lambda| \ll 1$,  and reads:
\begin{align}\label{rainvertNH}
\Gamma^{(3)} (p,k) \simeq \lambda \gamma^5~.
\end{align}
In \cite{alex2} and here, 
we restrict ourselves to the case  of real $\mu \in {\mathbb R}$. This stems from the fact that we are interested in the (physically relevant) case where the {\it energies} of the system in the massive phase are {\it real},  for which one must have the following condition among the (dynamically generated) mass parameters~\cite{qft2,alex1}
\begin{align}\label{mumrel}
|\mu| \le |m|.
\end{align}
The reality of the energy eigenspectrum in this case may be traced back~\cite{alex1,alex2} to the underlying $\mathcal C$PT (antilinear)symmetry of
the field theory \eqref{actionnhyuk}, with $\mathcal C$ appropriately defined (\eqref{cpt})~\cite{AMS}, according to the general arguments of \cite{most2,most3,most,maninnerPT,antilin}. 
As discussed in ~\cite{AB,AMS}, the condition \eqref{mumrel} guarantees unitarity, in the sense of the existence of a well-defined, {\it conserved},  probability density for the non-Hermitian fermionic system, which is less than unity.

Upon  assuming vanishing external momenta $k=0$ in the SD equations \eqref{SD1}, 
and performing the Euclidean momentum integrations with an UV cutoff $\Lambda$, we arrive at~\cite{alex1} 
\begin{align}
\label{sdnhb}
  M^2 =  \frac{\lambda^2}{4 \pi^2} \left[ \frac{(\Lambda^2 + m^2 - 3 \mu^2) \Lambda^2}{\Lambda^2 + m^2 - \mu^2} - (m^2 - 3 \mu^2) \ln \left( 1 + \frac{\Lambda^2}{m^2 - \mu^2} \right) \right]
\end{align}
\begin{align}
\label{sdnha}
  m + \mu \gamma^5 = - \frac{\lambda^2}{16 \pi^2} \frac{(m - \mu \gamma^5)}{M^2 - m^2 + \mu^2} \left[ M^2 \ln \left( 1 +
  \frac{\Lambda^2}{M^2} \right) - (m^2 - \mu^2) \ln \left( 1 + \frac{\Lambda^2}{m^2 - \mu^2} \right) \right]
\end{align}
Following our previous discussion in section \ref{sec:ener} on the energetics of the action \eqref{Sefffermion}, implying non-formation of a dynamical chiral mass $\mu \ne 0$, we set from now on $\mu=0$~\cite{alex2}.

On setting $m=\mu=0$, which is consistent with \eqref{mumrel}, we observe from \eqref{sdnhb} that there is now a consistent solution for the dynamically generated axion mass $M$, since \eqref{sdnhb} leads to
\begin{align}\label{scalar}
M^2 =  \frac{\lambda^2}{4 \pi^2} \Lambda^2 \, \ll \Lambda^2, \quad \lambda^2 \ll 1,
\end{align}
which is the same as the expression \eqref{dynam}. On account of \eqref{mumrel}, the case where $m =0$ but $\mu \ne 0$ is not allowed, as it would lead to unphysical situations with complex energies . 

For $\mu=0$, and $m \, M \ne 0$, we have the following system of SD equations
\begin{align}
\label{sd2mu0nh}
  M^2 =  \frac{\lambda^2}{4 \pi^2} \left[ \Lambda^2 - m^2 \ln \left( 1 + \frac{\Lambda^2}{m^2} \right) \right]
\end{align}
\begin{align}
\label{sd1mu0nh}
  1 = - \frac{\lambda^2}{16 \pi^2} \frac{1}{M^2 - m^2} \left[ M^2 \ln \left( 1 + \frac{\Lambda^2}{M^2} \right) - m^2 \ln \left( 1 + \frac{\Lambda^2}{m^2} \right) \right]
\end{align}

We will now consider partial solutions $0 \ne m \simeq M \ll \Lambda$. From (\ref{sd1mu0nh}):
\begin{align}
  - \frac{16 \pi^2}{\lambda^2} = \ln \left( 1 + \frac{\Lambda^2}{M^2} \right), 
\end{align}
which is inconsistent. It is also readily seen that the case 
$\mu=M=0$ also does not lead to fermion mass generation.

Hence dynamical fermion mass is not possible for pure anti-Hermitian Yukawa interactions, only scalar mass can be generated dynamically, in agreement with the generic energetics arguments~\cite{alex1}  provided in the previous subsection \ref{sec:ener}.

\subsection{Non-Hermitian Yukawa Interactions in the presence of attractive four-fermion interactions \label{sec:nh4f}}

\medskip

We next proceed to discuss explicitly the r\^ole on dynamical-mass generation played by additional {\it Hermitian attractive} four-fermion interactions in the non-Hermitian Yukawa model~\cite{alex2}, {\it i.e.} we consider the case $g=1$ in \eqref{actionnhyuk}. The pertinent Euclidean generating functional reads:
\begin{align}\label{part4f}
  Z^E [J, \eta, \overline{\eta}] &= \int \mathcal{D} [\phi  \psi
  \overline{\psi}] \exp \left\{ - \int d^4 x  \left[ \frac{1}{2} \, \partial_{\mu} \phi
  \partial^{\mu} \phi  + \overline{\psi} i\slashed{\partial} \psi - \lambda
  \phi \,  \overline{\psi} \gamma^5 \psi  + \frac{1}{2 f^2_4} (\overline{\psi} \, \gamma^5 \psi)^2\right] \right.
 \nonumber \\ &\left. - \int d^4 x [J  \phi  + \overline{\psi} \eta + \overline{\eta} \psi] \right\} \nonumber \\
& = \int \mathcal{D} [\sigma \phi \psi \overline{\psi}] \exp \left\{ - \int d^4 x \left[ \frac{1}{2} \partial_{\mu} \phi \partial^{\mu} \phi + \overline{\psi} i \slashed{\partial} \psi - \lambda \phi \overline{\psi} \gamma^5 \psi + \frac{f^2_4}{2} \sigma^2  + i \, \sigma \overline{\psi} \gamma^5 \psi \right] \right. \nonumber \\ &\left.- \int d^4 x [J \phi  + \overline{\psi} \eta + \overline{\eta} \psi + K \sigma] \right\},
\end{align}
where in the second equality we have linearised the four-fermion interactions using an auxiliary field $\sigma$. 
The additional ingredient with  respect to the analysis in the previous subsection is the $\sigma$ propagator, which in 
momentum space is given by $G_\sigma(p)=1/f_4^2$. 

As a result of the extra field $\sigma (x)$, the SD equations with vanishing external momenta read now~\cite{alex2}:
\begin{equation}
  G^{- 1}_f (0) - S^{- 1}_f (0) = - \lambda^2 \gamma^5 \left( \int \frac{d^4 p}{(2 \pi)^4} G_f (p) \gamma^5 G_s (p) \right) - \gamma^5 \left( \int \frac{d^4 p}{(2 \pi)^4} G_f (p) \gamma^5 G_{\sigma} (p) \right)
\end{equation}
\begin{equation}
  G^{- 1}_s (0) - S^{- 1}_s (0) =  \lambda^2 \tmop{tr} \left[ \gamma^5 \int \frac{d^4 p}{(2 \pi)^4} G_f (p) \gamma^5 G_f (p) \right]
\end{equation}
Using the dressed inverse propagators, as in the previous subsection, 
and performing the integrals using an UV cutoff $\Lambda$, we arrive at the following system of algebraic equations:
\begin{eqnarray}
  m + \mu \gamma^5 = - \frac{\lambda^2}{16 \pi^2} \frac{(m - \mu \gamma^5)}{M^2 - m^2 + \mu^2} \left[ M^2 \ln \left( 1 +
  \frac{\Lambda^2}{M^2} \right) - (m^2 - \mu^2) \ln \left( 1 + \frac{\Lambda^2}{m^2 - \mu^2} \right) \right] \nonumber\\
  + \frac{(m - \mu \gamma^5)}{16\pi^2 f_4^2} \left( \Lambda^2 - (m^2 - \mu^2) \ln \left( 1 + \frac{\Lambda^2}{m^2 - \mu^2} \right) \right)
\end{eqnarray}

\begin{equation}\label{massMnh}
  M^2 =  \frac{\lambda^2}{4 \pi^2} \left[ \frac{(\Lambda^2 + m^2 - 3 \mu^2) \Lambda^2}{\Lambda^2 + m^2 - \mu^2} - (m^2 - 3 \mu^2) \ln \left( 1 + \frac{\Lambda^2}{m^2 - \mu^2} \right) \right]
\end{equation}

As the SD pseudoscalar mass equation is independent of $f_4$, it is straightforward to see from \eqref{massMnh} that, for $m=\mu=0$, one obtains 
the dynamically generated (pseudo)scalar mass \eqref{scalar}.

If the case $\mu=0$ but $ m\,M \ne 0$, the system of SD equations reads
\begin{align}
\label{nh4mu0a}
  1 &= - \frac{\lambda^2}{16 \pi^2} \frac{1}{M^2 - m^2} \left[ M^2 \ln \left( 1 + \frac{\Lambda^2}{M^2} \right) - m^2 \ln \left( 1 + \frac{\Lambda^2}{m^2}\right) \right] + \frac{1}{16 \pi^2 f_4^2} \left( \Lambda^2 - m^2 \ln \left(1 + \frac{\Lambda^2}{m^2} \right) \right) \nonumber \\
M^2 &=  \frac{\lambda^2}{4 \pi^2} \left[ \Lambda^2 - m^2 \ln \left( 1 + \frac{\Lambda^2}{m^2} \right) \right], 
\end{align}
which for the specific case 
\be\label{meqM}
\Lambda \, \gg \, m \simeq M  \ne 0, 
\ee
leads to
 \begin{align}
\label{nh4mu0ab}
  1 =& - \frac{\lambda^2}{16 \pi^2}  \ln \left( 1 + \frac{\Lambda^2}{M^2} \right)  + \frac{m^2}{4\, f_4^2 \, \lambda^2}  \, \Rightarrow \, m^2 \simeq M^2 \simeq 4 \, \lambda^2 \, f_4^2  +  \Big | \mathcal O (\lambda^4 \, \ln \lambda^2) \Big |~,  
 \nonumber \\
  M^2 &\simeq m^2  \simeq   \frac{\lambda^2}{4 \pi^2} \Lambda^2~.
\end{align}
For consistency we then obtain
\be\label{4ff4}
f_4 \simeq \frac{\Lambda}{4\pi} -  \Big | \mathcal O (\lambda^2\, \ln \lambda^2) \Big |, \quad \lambda^2 \ll 1.
\ee
Thus, in non-Hermitian Yukawa interactions, upon the inclusion of sufficiently strong four-fermion attractive interactions, one can obtain dynamical mass for both fermions and axions. 

Some important remarks are in order at this point. The reader should have noticed from 
\eqref{nh4mu0ab} that there is SD dynamical mass for {\it arbitrarily small values} of the Yukawa coupling $\lambda$, which thus includes the case where such couplings are generated by stringy instanton effects, as discussed in section \ref{sec:modelaxion} ({\it cf.} \eqref{instl}). 
We stress that this is a consequence of the fact that the prototype model does not contain self-interactions of the 
axion-$\phi$ field, which is motivated by the stringy considerations of section \ref{sec:modelaxion} ({\it cf.} discussion following \eqref{pformaxion}). We note that, in the presence of such self-interactions, {\it eg.} a $\phi^4$-term, the situation concerning fermion mass generation can be very different. For instance, in the specific (Hermitian) model of \cite{bashir}, where there is a common coupling between the Yukawa and $\phi^4$ self-interactions, there is a critical value of $g$ above which dynamical fermion mass is possible. We do not consider axion self-interactions throughout this work or in \cite{alex1,alex2,soto}.

We also note that for our purposes it suffices that we considered the special case \eqref{meqM} for the dynamical masses of axions and fermions of approximately equal magnitude. More general solutions of course may exist, but they are not relevant for our purposes, which are to demonstrate explicitly a non-trivial concrete solution of  the SD system entailing dynamical masses for both axions and fermions.

We remark at this point that, although above we view the interaction \eqref{4feff} as an effective attractive interaction
which incorporates the subtraction of repulsive terms of similar form stemming from the quantum-torsion-induced repulsive interactions \eqref{sea6II} upon applying the Fierz identities \eqref{fierz}, nonetheless one may also consider the case in which this interaction genuinely dominates the torsion terms, for an appropriate value of the cut-off scale $\Lambda$, which defines the momentum scale above which the validity of the low-energy effective field theory breaks down. In such a case one would justify ignoring the repulsive interactions altogether. To determine in this case the scale $\Lambda$, we should 
compare the coefficient $1/(2f_4^2)$ of the attractive terms \eqref{4feff}, with $f_4$ determined by \eqref{4ff4}, with the corresponding coupling $\sqrt{3\kappa^2/8}$ of the repulsive quantum-torsion-induced terms \eqref{fierz}. 
The reader observes that 
dominance of the attractive term \eqref{4feff} by, say, at least one order of magnitude, over the corresponding term in \eqref{fierz},  would imply a cut-off momentum scale  
\be\label{fierzcutoff}
\frac{1}{2\, f_4^2} \gtrsim {\mathcal O}(10) \times \frac{3\, \kappa^2}{8}  \quad \stackrel{{\rm cf.}\,\eqref{4ff4}}{\Rightarrow} 
\quad \Lambda \lesssim {\mathcal O}\Big(\sqrt{\frac{4\,\pi}{15}}\Big) \, M_P \sim 0.9 \, M_P~ ,
\ee which is phenomenologically reasonable for an effective theory embedded in an UV complete quantum gravity theory, such as strings. In fact, the range for $\Lambda$ in \eqref{fierzcutoff} is compatible with \eqref{alphakappa}, 
which is satisfied for 
\be\label{LamMs}
\Lambda  \sim M_s \simeq 0.2 \, M_P~, 
\ee
implying that $M_s$ acts as an effective cut-off energy scale above which the validity of the point-like string-effective theory breaks down. Under \eqref{LamMs}, the dynamically generated mass \eqref{nh4mu0ab} becomes 
\be\label{dmax}
M \simeq m  \simeq   \frac{|\lambda|}{2 \pi} \, \Lambda \simeq 0.03\, |\lambda| \, M_P \simeq 0.04\, |\lambda|  \times 10^{19}~{\rm GeV}.
\ee
If $\lambda$ is generated by string instantons \eqref{instl}, then, in order to have axion masses larger than $10^{-24}$~eV, which is a lower-bound for the mass of an axion in case the latter plays the r\^ole of a dominant dark-matter species~\cite{marsh}, one needs the (real parts of the) stringy-instanton actions to be of order:
$0 < S_{{\rm instanton}} < 116.5  $ (in natural units of $\hbar c=1$). Smaller axion masses $> 10^{-32}$~eV are allowed if the axion constitutes only a small ({\it e.g.} $ < 5\%$) part of dark matter in the Universe, in which case 
one obtains for the instanton actions $0 < S_{{\rm instanton}} < 134.9.$ 
Such considerations allow for a rather wide variety of axion masses in our scenarios, which can thus play a r\^ole as dark matter candidates, including ultralight ones, with sensitivity to be falsified at immediate-future experimental facilities.\footnote{Such instanton actions are smaller in magnitude than the ones required for axion quintessence in Hermitian stringy axion cosmologies, in which the contributions of axions to dark energy are non-negligible; in such cases one needs $S_{\rm instanton} \sim 200-300$~\cite{marsh}.  We remark at this point, however, that, strictly speaking, one cannot directly compare the  phenomenological constraints discussed in the context of our non-Hermitian models with the phenomenological constraints for generic Hermitian stringy axion models based on instanton-induced axion potentials~\cite{witten}, as a consequence of the anomalous couplings of the (stringy) axions. The dynamical axion masses \eqref{dmax} pertain to a model \eqref{actionnhyuk} without anomalous terms ($\gamma=0$) and no potential for the pseudoscalar fields. Although non-Hermitian anomaly terms ($\gamma \ne 0$)  are discussed in the next section \ref{sec:anom}, nonetheless for such non-Hermitian  models the concept of instanton-induced potentials is not completely understood, as already mentioned. Hence, the considerations in \cite{witten} might not apply even to that case.} 

We also note that the presence of Yukawa interactions, coupling the axion to right-handed neutrinos in the model of \cite{mp}, under the assumption \eqref{meqM}, \eqref{nh4mu0ab} for the dynamically generated masses, can be reconciled with phenomenologically realistic situations of particle physics by considering such ultralight right-handed neutrinos as being almost decoupled from the standard-model matter sector. In other words, the dominant coupling of such right-handed neutrinos with masses given by \eqref{nh4mu0ab} to matter is the Yukawa one, while their mixing through Higgs-portal-type interactions with standard-model matter is negligible. Of course, such restrictions are lifted if one views the prototype models discussed here and in \cite{alex1,alex2,soto} as purely phenomenological, not necessarily connected to microscopic string models, in which case  $\lambda$ is a free parameter, which need not be generated by string instanton effects, and hence need not be that small.

\subsection{Resummation effects of strong four-fermion interactions \label{sec:4f}} 
\medskip

\begin{figure}[t]
 \centering
  \includegraphics[clip,width=0.7\textwidth]{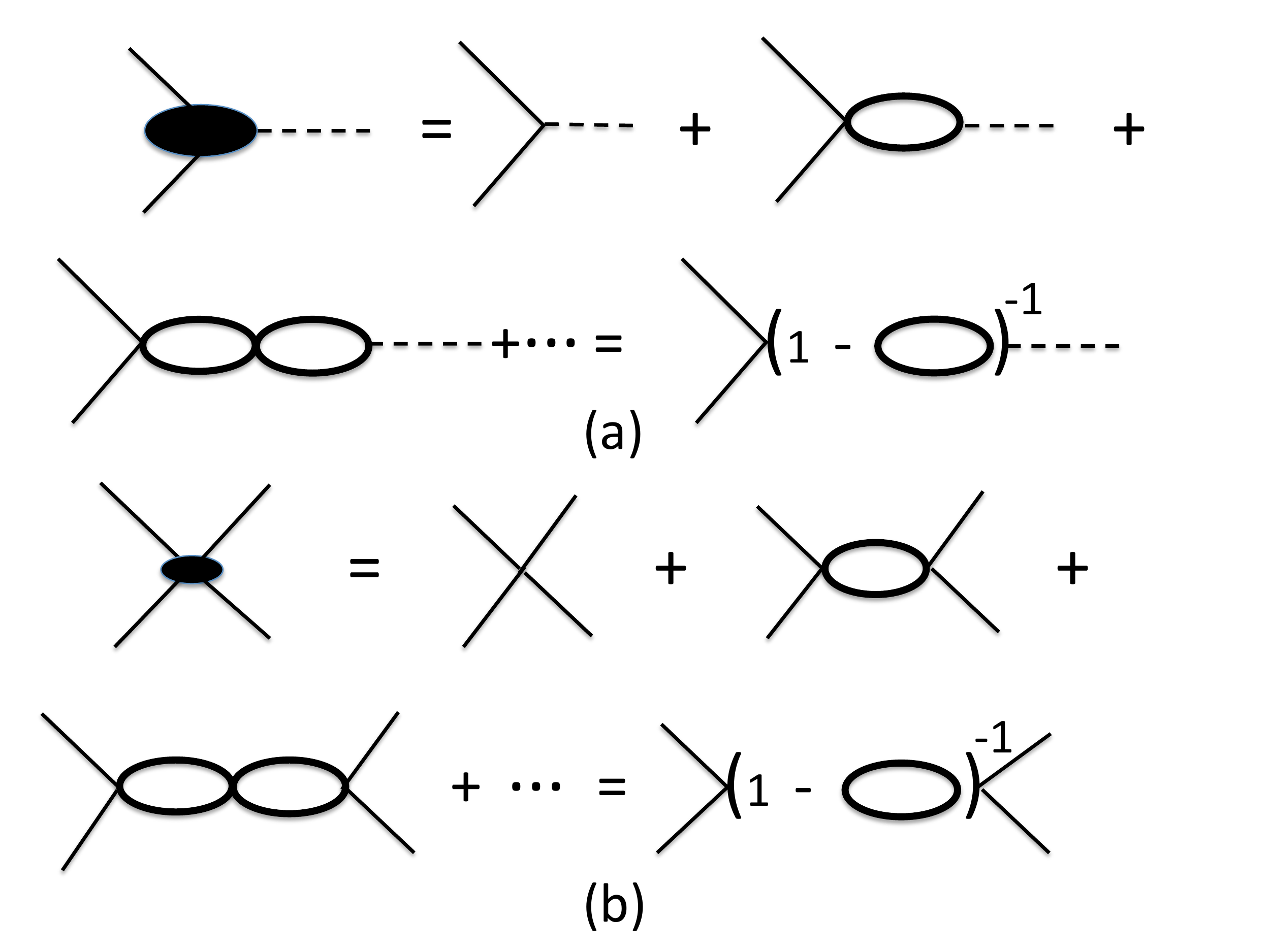} 
\caption{\it Resummation (dark elliptical blobs) of leading (quadratic) UV  divergences due to four-fermion `bubble' diagrams (ellipse without filling in the figures). \underline{Upper figure {\rm (a)}}: effects on Yukawa interaction vertex.
\underline{Lower figure {\rm  (b)}}: effects on four-fermion vertex. Continuous lines refer to fermions, dashed lines pertain to pseudoscalar fields. Figures taken from \cite{alex2}. }\label{fig:diagr}
\end{figure}
Before closing this section we would like to comment briefly on how resummation of the four-fermion interactions
affects qualitatively the above conclusions. Indeed, this is an important aspect, given the strong coupling nature of the
dimensionless four-fermion coupling \eqref{4ff4}
\be 
\frac{\Lambda}{\sqrt{2}\, f_4} \sim 4\pi  > 1,
\ee
which implies the necessity for a resummations of the loop ``bubble'' diagrams associated with quantum corrections of the four-fermion vertex, as indicated in fig.~\ref{fig:diagr}.
This issue was discussed in detail in \cite{alex2}, where we refer the interested reader for details. For our purposes here, we only mention that the result of such a resummation, as far as the leading quadratic UV divergences are concerned, of interest to us here, is that the relevant quantum corrections can be absorbed in ``renormalised'' Yukawa and four-fermion couplings: 
\be\label{renormcoupl}
 \quad f_{4R} \sim \frac{1}{\sqrt{2}}\, f_4~, \quad \lambda_R \sim 2\, \lambda~,
\ee
which are thus of the same order as the bare couplings. 
Therefore, the validity of the above SD analysis and the pertinent conclusions are guaranteed upon replacing the bare couplings $\lambda, f_4$ by the `renormalised' ones \eqref{renormcoupl}~\cite{alex2}. 
This is to be understood in all the previous formulae as well as the ones that follow. 
For notational convenience, we omit from now on the suffix $R$.

\section{Inclusion of non-Hermitian anomaly terms and non-Hermitian axial backgrounds \label{sec:anom}}

\medskip

A final topic we would like to discuss in this mini-review is the r\^ole of {\it non-hermitian anomaly and axial background} terms on the dynamical mass generation induced by the non-Hermitian Yukawa interactions in the model. The presence of the anomalous terms have been responsible, as we have discussed in section \ref{sec:model}, for the radiative generation of Majorana masses for the right-handed neutrinos in the Hermitian model of \cite{mp}. Formally, when both Yukawa  and anomalous interactions  are characterised by purely imaginary couplings \eqref{analytically}, {\it i.e.} when such interactions are anti-Hermitian, the expressions for the radiative fermion masses  \eqref{massR2} retain their reality. However, the concept of anomalies, and the relevant studies, are not yet  developed for non-Hermitian theories, except making formal analogies, and therefore it would be interesting to examine alternative ways for fermion (and axion) mass generation, through the anti-Hermitian Yukawa interaction, and scrutinise the r\^ole of the presence of anomalous terms on such a phenomenon. This has been done in detail in \cite{soto}. Below we shall only review the basic conclusions of that work.

In addition, we shall also examine in this section the r\^ole of non-Hermitian interactions of fermions with a constant axial background $\mathcal B_\mu$ 
\be\label{axial}
\mathcal S^{antiherm}_{\rm axial-backgr} = \int d^4 x \sqrt{-g} \, i\, \mathcal B_\mu \overline \psi\, \gamma^\mu \, \gamma^5 \, \psi ~, \quad \mathcal B_\mu \in \mathbb R~.
\ee
In the context of microscopic string theories, discussed in section \ref{sec:model}, the microscopic origin of the non-Hermitian background can be traced back to the non-Hermitian anomalous interactions of the KR gravitational axion of the massless bosonic string multiplet in some formulations of the theory ( see ``{\bf Scheme II}'', \eqref{sea6II}).  To understand this, we mention that, in Hermitian theories, i.e. when $i\mathcal B_\mu \in \mathbb  R$, such constant backgrounds (with temporal components only in a given Lorentz frame, specifically the cosmological Robertson-Walker (RW) frame) have been argued to characterise cosmological configurations of the KR gravitational axion field, $b(x)$, exhibiting a linear dependence in the cosmic time, $t$, in the RW frame~\cite{decesare,anomalies},
\be\label{KRaxionlinear}
b(x) = ({\rm constant}) \times  t~.
\ee
Non-hermitian constant axial backgrounds \eqref{axial}, then, can arise in the ``{\bf Scheme II}'' \eqref{sea6II} of the effective string theory actions, in the way discussed in section \ref{sec:model}. 
Moreover, in Hermitian models, constant axial backgrounds 
provide  alternative ways for tree-level Leptogenesis~\cite{decesare,boss} ({\it i.e.} the generation of Lepton-number asymmetry between particles and anti-particles), in theories involving massive right-handed neutrinos, as a result of the asymmetric decays of the latter to standard-model leptons and antileptons in the presence of such backgrounds. It is yet to be seen whether non-Hermitian axial backgrounds \eqref{axial} can lead to similar Leptogenesis phenomena in non-Hermitian theories. 

Their contribution to mass generation, though, is one such aspect that can be settled already. In the work of \cite{soto}, which will be reviewed briefly here, we shall study the r\^ole of such interactions on the masses of axions and fermions,  generated dynamically by non-Hermitian Yukawa couplings. We also note that in \cite{NJLnh} such non-Hermitian constant backgrounds have been examined from the point of view of their r\^ole in fermion mass generation in the context of Hermitian Nambu-Jona-Lasinio four-fermion models~\cite{NJL}, which however did not contain Yukawa interactions, par contrast to our case here, where the latter constitute the dominant source for mass generation. 

Therefore, in this section we shall study the following prototype action:
{\small \begin{eqnarray}\label{bacanomlag}
 \mathcal{S}^{anti-Herm}_{axial-back+anom} &=&  \int d^4 x \sqrt{-g} \, \Big(\frac{1}{2} \partial_{\mu} \phi \partial^{\mu} \phi + \bar{\psi} i \slashed{\partial} \psi - \frac{i\,\gamma}{f_b}
  (\partial_{\mu} \phi) \bar{\psi} \gamma^{\mu} \gamma^5 \psi \nonumber \\ 
  &+&  i\,\mathcal{B}_{\mu} \bar{\psi}
  \gamma^{\mu} \gamma^5 \psi 
   + \lambda \, \phi \,  \bar{\psi} \gamma^5 \psi - \frac{g^2}{2 f^2_b}
  (\bar{\psi} \gamma^5 \psi)^2\,\Big).
\end{eqnarray}}We remark at this stage that, in the context of the model \cite{mp}, $f_b$  in \eqref{bacanomlag} is given by \eqref{fb}, but here we consider it  as an arbitrary mass scale to be determined self consistently in terms of the UV cutoff $\Lambda$ in the phase where dynamical mass generation occurs~\cite{soto}.  

In \eqref{bacanomlag} we have included non-trivial four-fermion attractive interactions, since, from arguments given previously (section \ref{sec:ener}), we know that their absence in anti-Hermitian Yukawa models disfavours dynamical mass generation for fermions. The coupling $g$ is inserted to demonstrate the difference (in general) of the  strength of the four-fermion attractive interactions from that of the anomalous terms. 
The anomalies are encoded in the four-divergence of the axial current ({\it cf.} \eqref{axialdiv}), and appear after partial integration of the relevant (third) term on the right-hand-side of \eqref{bacanomlag}, with coefficient  $-i\,\gamma/f_b$. In what follows, consistently with the discussion so far, we shall work in a model formulated in a Minkowski spacetime. 

The action \eqref{bacanomlag} is $\mathcal C PT$ invariant, under the transformations \eqref{cpt} for the fermions, which guarantees the reality of the energy spectrum, and hence the dynamical masses~\cite{soto}, according to the generic analysis of \cite{most2,most3,most,maninnerPT,antilin}. We note that a constant anti-Hermitian axial background would break {\it spontaneously} the (complex) Lorentz symmetry, and thus the generalised $\mathcal C$PT symmetry, as in  the Hermitian case~\cite{decesare,boss}. Nonetheless, the reality of the dynamically-generated masses is guaranteed~\cite{soto}, as the condition \eqref{mumrel} is maintained.

\subsection{Models with non-Hermitian Yukawa and anomalous interactions - no axial background \label{ref:nhyukanom}}

\medskip

We commence our study with the case of anti-Hermitian Yukawa interactions in the presence of anti-Hermitian 
anomalous couplings to fermions, in the absence of axial background, i.e. setting $\mathcal B_\mu=0$ in \eqref{bacanomlag}.  In this case, the Euclidean partition function, under the linearization of the four-fermion interactions by means of the auxiliary scalar $\sigma$, reads (upon the standard inclusion of appropriate sources
$\eta, \bar \eta, K, J$ for the fields $\psi$, $\overline \psi$, $\sigma$ and $\phi$, repsectively):
\begin{eqnarray}\label{zeff}
  Z [K, J, \eta, \bar{\eta}] &=& \int \mathcal{D} [\sigma \phi \psi \bar{\psi}]
  \exp \Big( - \int d^4 x \Big[ \frac{1}{2} \partial_{\mu} \phi \partial^{\mu}
  \phi + \bar{\psi} i \slashed{\partial} \psi  \nonumber \\  &+&  i \frac{\gamma}{f_b}
  (\partial_{\mu} \phi) \bar{\psi} \gamma^{\mu} \gamma^5 \psi - \lambda\,
  \phi \, \bar{\psi} \gamma^5 \psi + \frac{f^2_b}{2 g^2} \sigma^2 - i \sigma
  \bar{\psi} \gamma^5 \psi \Big] \nonumber\\
   & &  - \int d^4 x [J \phi + \bar{\psi} \eta +
  \bar{\eta} \psi + K \sigma] \Big)
  \end{eqnarray}
In the way written, the anomalous interactions with coupling $\gamma$ in \eqref{bacanomlag} imply 
that the Feynman rule for the (bare) axion-fermion vertex $\phi\, \bar{\psi}\psi$ is given by $\left(\lambda - \frac{\gamma}{f_b} \slashed{p} \right)\gamma^5$, in the convention where all the momenta are incoming to the vertex.

The details of the SD analysis are given in \cite{soto} and will not be repeated here, where we only state the final result.
We seek a partial solution of the SD equations for dynamical mass generation stemming from \eqref{zeff}, 
for which there is an approximate equality of fermion ($m$) and axion ($M$) dynamical masses, $m \simeq M$,  with the chiral mass $\mu=0$,  which suffices for our purposes, for reasons explained above.
Using the rainbow approximation for the vertices, appropriate for small Yukawa and anomalous couplings 
$|\gamma| \sim |\lambda | \ll 1$, assumed here, and  taking into account the effects of resummation of the four-fermion interactions, as described in the previous section \ref{sec:4f}, we arrive at the following system of SD equations
describing the dynamical generation of fermion and pseudoscalar masses~\cite{soto}

\begin{eqnarray}
\label{sol1nohermb}
  1 &=& - \frac{\lambda^2}{16 \pi^2} \frac{1}{M^2 - m^2} \left[
  M^2 \ln \left( 1 + \frac{\Lambda^2}{M^2} \right) - m^2 \ln \left( 1 +
  \frac{\Lambda^2}{m^2} \right) \right] \nonumber\\
  & & + \frac{\gamma^2}{16
  \pi^2 f^2_b} \frac{1}{M^2 - m^2} \left[ - \Lambda^2 (M^2 - m^2) + M^4 \ln
  \left( 1 + \frac{\Lambda^2}{M^2} \right) - m^4 \ln \left( 1 +
  \frac{\Lambda^2}{m^2} \right) \right] \nonumber\\
  & & + \frac{g^2}{16 \pi^2 f^2_b} \left(
  \Lambda^2 - m^2 \ln \left( 1 + \frac{\Lambda^2}{m^2} \right) \right)
\end{eqnarray}
and
\begin{equation}
\label{sol2nohermb}
  M^2 = \frac{\lambda^2}{4 \pi^2} \left[ \Lambda^2 - m^2 \ln
  \left( 1 + \frac{\Lambda^2}{m^2} \right) \right],
\end{equation}
where $\Lambda$ is the UV cut-off, as in previous sections. 
Seeking solutions with $m \simeq M \ne 0$ ({\it cf.} (\eqref{meqM}) we obtain
\begin{equation}
\label{nhanf}
  1 = - \frac{1}{16 \pi^2} \left( \lambda^2 -
  \frac{\gamma^2 M^2}{f^2_b} \right) \ln \left( 1 +
  \frac{\Lambda^2}{M^2} \right) + \frac{g^2-\gamma^2}{16
  \pi^2 f^2_b} \left[ \Lambda^2 - m^2 \ln \left( 1 + \frac{\Lambda^2}{M^2}
  \right) \right]
\end{equation}
and
\begin{equation}
\label{nhans}
\frac{4 \pi^2}{\lambda^2} M^2 = \left[ \Lambda^2 - m^2 \ln
\left( 1 + \frac{\Lambda^2}{m^2} \right) \right]
\end{equation}

Considering $\Lambda^2\gg M^2$, $|\lambda|, |\gamma| \ll 1$,  with $g \ne \gamma$,  and substituting \eqref{nhans} in \eqref{nhanf}, we arrive at~\cite{soto}:
\begin{equation}
\label{nhmas}
M^2\simeq m^2 \simeq \frac{4 f^2_b \lambda^2}{g^2-\gamma^2} + | \mathcal{O} \Big((\lambda^4, \lambda^2 \gamma^2)\, \ln(\lambda^2)\Big) |
\end{equation}

Hence, consistency between \eqref{nhmas} and \eqref{nhans} requires 

\begin{equation}\label{dmass}
\frac{f_b}{\sqrt{g^2-\gamma^2}} \simeq \frac{\Lambda}{4
\pi} - | \mathcal{O} (\lambda^2) | \quad \Rightarrow \quad M^2 \simeq m^2 \simeq \frac{\lambda^2}{4\pi^2}\, \Lambda^2 +  | \mathcal{O} \Big((\lambda^4, \lambda^2 \gamma^2)\, \ln(\lambda^2)\Big) |\,,
\end{equation}
for small $\gamma, \lambda $ couplings. In order to have dynamical mass generation one must have\footnote{For a discussion on the limiting case  $g^2=\gamma^2 + \epsilon$, $\epsilon\, \rightarrow \, 0^+$, we refer the reader to \cite{soto}.} 
\be\label{restr}
g^2>\gamma^2, 
\ee
We remark that the form \eqref{dmass} of the dynamical mass in terms of the UV cutoff $\Lambda$ is similar to the 
case \eqref{nh4mu0ab}, although the reader should bear in mind the connection between $\Lambda$ and $f_b$ in this case, \eqref{nhmas}, which depends on the strength of the anomalous coupling. In this respect, the phenomenological analysis related to \eqref{dmax} in section~\ref{sec:nh4f}, would also apply in this case. The magnitude of the parameter $\gamma$, entering the anomalous interactions, is treated phenomenologically, although, as already mentioned, in string-inspired models~\cite{mp}, the latter is related to the kinetic mixing between model-dependent and model-independent(KR) axions in string theory, \eqref{mixing}, and it could also be viewed, in some cases, as the result of non-perturbative effects, similarly to the (non-perturbatively-induced) Yukawa-coupling interactions \eqref{yuk}, \eqref{instl}.  

The reader should also have noticed that bare axion and fermion masses are not necessary, which implies that dynamical mass generation is possible under the above restriction \eqref{restr}.
Thus in this case, contrary to the Hermitian one (which is studied in \cite{soto}), it is a sufficiently strong attractive 
four-fermion interaction that catalyses the Yukawa-induced dynamical mass generation, since the non-Hermitian anomaly term {\it resists} mass generation. 

In \cite{alex1,soto} the case of Majorana fermions has also been studied, which are of relevance to the model of \cite{mp}. The case is qualitatively similar to 
the Dirac case studied above, apart from some numerical factors in the expression for the dynamical mass and the one providing the connection of the four-fermion couplings $f_b$ to the cut-off $\Lambda$. 
In such a case, the analogue of \eqref{nhmas} reads: 
\be\label{majorana}
 M^2_{\rm Dyn~Maj} \simeq \frac{2 f_b^2\, \lambda^2}{g^2-\gamma^2} + \dots, \quad {\rm with} \quad 
\frac{f_b}{\sqrt{g^2-\gamma^2}} \simeq \frac{\Lambda}{2\sqrt{2} \pi},
\ee 
 implying:
\begin{align}\label{massR2comp}
M_{\rm Dyn~Maj}^{anti-herm} \sim  \frac{\lambda }{2 \pi} \, \Lambda, \quad |\lambda| \ll 1. 
\end{align}
Upon comparison of \eqref{massR2comp} with \eqref{massR2} (with $ |\gamma| \, , \, |\lambda |\ll 1 $, as required for consistency of our SD treatment, which also characterises the physically motivated case in which these interactions are generated by non-perturbative stringy effects, as discussed in section \ref{sec:model}), one obtains:
\begin{align}\label{massR2compb}
M_R^{\rm anti-herm} \, \sim \,  \frac{\sqrt{3/2}\, \gamma\  (\kappa \, \Lambda)^5}{49152 \,
\pi^3}\, M_{\rm Dyn~Maj}^{\rm anti-herm} \simeq 8 \cdot 10^{-7} \, \gamma \, (\kappa \, \Lambda)^5 
\, M_{\rm Dyn~Maj}^{\rm anti-herm} \ll M_{\rm Dyn~Maj}^{\rm anti-herm}\; ,
\end{align}
for any value of a sub-planckian cutoff $\kappa \, \Lambda \,  \lesssim 1$. Hence in the non-Hermitian case, the anomaly-and-Yukawa-induced dynamical mass dominates the radiative anomalously-induced mass, provided the existence of the latter is rigorously proven of course in this case, something which is currently pending.

\subsection{Non-Hermitian axial background and Yukawa interactions - no anomalies \label{sec:axialbnanom}}

\medskip 

As a final topic for discussion in this review, we consider the case where an anti-Hermitian axial background \eqref{axial} is present for an anomaly-free fermion model, {\it i.e.} upon setting $\gamma=0$ in \eqref{bacanomlag}. 
The incorporation of both non-Hermitian axial background and anomalous interactions is straightforward and presents no conceptual difficulties, only algebraically more elaborate analysis. Hence it will not be discussed here. 

For details of the SD study of models with both anti-Hermitian background and Yukawa interactions we refer the reader to \cite{soto}. Below we shall only give the final results. The pertinent solutions for SD mass generation  for axions and fermions (with  approximately equal masses $m\simeq M ({\it cf.} $\eqref{meqM}) and chiral mass $\mu=0$) read:

{\small \begin{eqnarray}
\label{scalnhbnhy}
1 &=& -\frac{\lambda^2}{16 \pi^2 \bar{M}^2 \bar{\mathcal{B}}^2}
\left( 1 + 2 (\bar{M}^2 + \bar{\mathcal{B}}^2) - \sqrt{(1 + \bar{M}^2 -
\bar{\mathcal{B}}^2)^2 + 4 \bar{M}^2 \bar{\mathcal{B}}^2} \right.\nonumber\\
& & \left. - \left( \sqrt{(1 +
\bar{M}^2 - \bar{\mathcal{B}}^2)^2 + 4 \bar{M}^2 \bar{\mathcal{B}}^2} -
(\bar{M}^2 + \bar{\mathcal{B}}^2) \right) (\bar{M}^2 + 7 \bar{\mathcal{B}}^2)\right.\nonumber\\
& &\left. + 4 \bar{\mathcal{B}}^2 (\bar{M}^2 - 2 \bar{\mathcal{B}}^2) \log \left[
\frac{1 + \bar{M}^2 - \bar{\mathcal{B}}^2 + \sqrt{(1 + \bar{M}^2 -
\bar{\mathcal{B}}^2)^2 + 4 \bar{M}^2 \bar{\mathcal{B}}^2}}{2 \bar{M}^2}
\right] \right).
\end{eqnarray}}
and 
{\small \begin{eqnarray}
\label{fermnhbnhy}
  1 &=& \frac{\lambda^2}{32 \pi^2 \bar{\mathcal{B}}^2} \left( 1
  + \bar{\mathcal{B}}^2 \left( 1 -
  \frac{\bar{\mathcal{B}}^2}{\sqrt{\bar{\mathcal{B}}^4 + 4 \bar{M}^2
  \bar{\mathcal{B}}^2}} \right) \log \left( 1 + \frac{1}{\bar{M}^2} \right) -
  \left( \sqrt{(1 + \bar{M}^2 - \bar{\mathcal{B}}^2)^2 + 4 \bar{M}^2
  \bar{\mathcal{B}}^2} - (\bar{M}^2 + \bar{\mathcal{B}}^2) \right) \right.\nonumber\\
  & & \left. - 3 \bar{\mathcal{B}}^2 \log \left[ \frac{1 + (\bar{M}^2 -
  \bar{\mathcal{B}}^2) + \sqrt{(1 + \bar{M}^2 - \bar{\mathcal{B}}^2)^2 + 4
  \bar{M}^2 \bar{\mathcal{B}}^2}}{2\bar{M}^2} \right] \right.\nonumber\\ 
  & & \left. + \frac{\bar{\mathcal{B}}^4}{\sqrt{\bar{\mathcal{B}}^4 + 4 \bar{M}^2
  \bar{\mathcal{B}}^2}} \log \left[ \frac{\left( \sqrt{\bar{\mathcal{B}}^4 + 4
  \bar{M}^2 \bar{\mathcal{B}}^2} + \sqrt{(1 + \bar{M}^2 -
  \bar{\mathcal{B}}^2)^2 + 4 \bar{M}^2 \bar{\mathcal{B}}^2} \right)^2 - (1 +
  \bar{M}^2)^2}{\left( (\bar{M}^2 + \bar{\mathcal{B}}^2) +
  \sqrt{\bar{\mathcal{B}}^4 + 4 \bar{M}^2 \bar{\mathcal{B}}^2} \right)^2 -
  \bar{M}^4} \right] \right) \nonumber\\
  & & - \frac{\bar{g}^2}{64 \pi^2
  \bar{\mathcal{B}}^2} \left( 1 + 2 (\bar{M}^2 + \bar{\mathcal{B}}^2) -
  \sqrt{(1 + \bar{M}^2 - \bar{\mathcal{B}}^2)^2 + 4 \bar{M}^2
  \bar{\mathcal{B}}^2} \right.\nonumber\\
  & & \left. - \left( \sqrt{(1 + \bar{M}^2 - \bar{\mathcal{B}}^2)^2
  + 4 \bar{M}^2 \bar{\mathcal{B}}^2} - (\bar{M}^2 +
  \bar{\mathcal{B}}^2) \right) (\bar{M}^2 + 7 \bar{\mathcal{B}}^2) \right.\nonumber\\
  & & \left. + 4 \bar{\mathcal{B}}^2 (\bar{M}^2 - 2 \bar{\mathcal{B}}^2) \log \left[ \frac{1
  + (\bar{M}^2 - \bar{\mathcal{B}}^2) + \sqrt{(1 + \bar{M}^2 -
  \bar{\mathcal{B}}^2)^2 + 4 \bar{M}^2 \bar{\mathcal{B}}^2}}{2\bar{M}^2} \right]
  \right),
\end{eqnarray}}where $\bar{M}=M/\Lambda$ and $\bar{\mathcal{B}}=\mathcal{B}/\Lambda$ and $\bar{g}=(g\Lambda)/f_b$. 

\begin{figure}[ht]
 \centering
  \includegraphics[clip,width=0.65\textwidth,height=0.25\textheight]{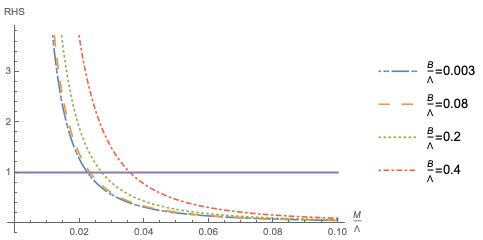} 
\caption{\it Plot of the right-hand side of \eqref{scalnhbnhy} versus the dynamical axion mass $M$ in units of the cut-off $\Lambda$, for the fixed value $\lambda^2=0.02$ for definiteness.  
All curves intersect the solid line, which represents the left hand side of \eqref{scalnhbnhy}, implying dynamical mass generation for the pseudoscalar field, without the need to introduce a bare mass for it. Figure taken from \cite{soto}.}\label{fig:plotscalnhbnhy}
\end{figure}

Figure \ref{fig:plotscalnhbnhy} plots the right hand side of \eqref{scalnhbnhy} as a function of the axion mass $\bar M$ for different values of $\mathcal{B}$. All the curves intersect the solid line, corresponding to the fixed value 1, which represents the left hand side of \eqref{scalnhbnhy}. The reader should observe that, for a non-Hermitian axial background, the larger the background, the bigger the dynamical mass. This is to be contrasted with the case of a Hermitian axial  background, for which, as we have discussed in \cite{soto}, one faces the opposite situation, that is, the larger the background, the smaller the dynamical mass. 

\begin{figure}[ht]
 \centering
  \includegraphics[clip,width=0.65\textwidth,height=0.25\textheight]{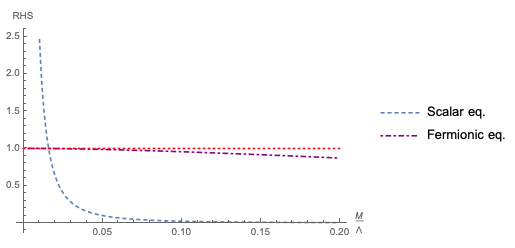} 
\caption{\it Plot of the right-had-side of \eqref{scalnhbnhy} (dashed line), for $\lambda^2=0.01$ and $\bar{\mathcal{B}}=0.0004$. The dotted-dashed line represents the right hand side of \eqref{fermnhbnhy} for the above values of $\lambda^2$ and $\bar{\mathcal{B}}$, and $\bar{g}=12.58$. The constant dotted line at 1 represents the left hand side of the equations \eqref{scalnhbnhy} and \eqref{fermnhbnhy}. The existence of a common intersection point for all three  curves demonstrates the existence  of dynamical mass generation, of approximately equal magnitude, for both fermions and axions. Figure taken from \cite{soto}.}\label{fig:plotnhbackferynh}
\end{figure}

From \eqref{scalnhbnhy} one obtains a value for $M$, which, upon being inserted in \eqref{fermnhbnhy}, determines the value of the four-fermion coupling $g/f_b$ for which there is a consistent dynamical mass for the fermion, of  approximately equal magnitude to the pseudoscalar mass, \eqref{meqM}. For concrete examples, we refer the reader to figure~\ref{fig:plotnhbackferynh}. We observe from the figure that the dashed line, corresponding to the right hand side of \eqref{scalnhbnhy}, intersects the constant dotted line at 1, representing the left-hand side of the equation. This implies the existence of a non-trivial solution for $M/\Lambda=0.159$. Using this value in \eqref{fermnhbnhy}, we then obtain a consistent solution for $\bar{g}=12.58$, in the sense that for this value of $\bar{g}$ the three curves have a common intersection at $M/\Lambda=0.159$.
This demonstrates that there is dynamical mass generation with $m \simeq M < \Lambda$ in this case for strong (dimensionless) four-fermion couplings $\bar{g} > 1$.

Although in our analysis above, and in \cite{alex1,soto} we restricted ourselves to the partial solution $m \simeq M$, we remark that solutions beyond this restriction also exist. For instance, if we
consider the case in which the Yukawa interactions are absent $\lambda^2\to 0$, and only the (attractive) four-fermion interaction is present, $g \ne 0$, in \eqref{bacanomlag}, then 
we recover the result of \cite{NJLnh} for dynamical mass generation only for fermions, for which the fermionic 
SD equation  reads~\cite{soto}
\begin{eqnarray}
\label{nhbackferm}
1 &=& - \frac{g^2}{64 \pi^2 f^2_b \mathcal{B}^2} \left( \Lambda^4 + \Lambda^2
\left( 2 (m^2 +\mathcal{B}^2) - \sqrt{(\Lambda^2 + m^2 -\mathcal{B}^2)^2 + 4
m^2 \mathcal{B}^2} \right) \right.\nonumber\\
& & \left. - (m^2 + 7\mathcal{B}^2) \left( \sqrt{(\Lambda^2 +
m^2 -\mathcal{B}^2)^2 + 4 m^2 \mathcal{B}^2} - (m^2 +\mathcal{B}^2) \right) \right.\nonumber\\
& & \left. + 4\mathcal{B}^2 (m^2 - 2\mathcal{B}^2) \log \left[ \frac{(m^2 -\mathcal{B}^2) +
\Lambda^2 + \sqrt{(\Lambda^2 + m^2 -\mathcal{B}^2)^2 + 4 m^2 \mathcal{B}^2}}{2
m^2} \right] \right).
\end{eqnarray}
This is essentially the same  as equation (34) in ref.~\cite{NJLnh}, where the Nambu Jona Lasinio  four-fermion model was studied in the presence of a constant  non-Hermitian axial background. The study of \eqref{nhbackferm} indicates~\cite{NJLnh} that the inclusion of the background increases the dynamical fermion mass due to the strong four-fermion interactions alone.

\section{Conclusions and Outlook \label{sec:concl}}

\medskip

In this mini-review we discussed some aspects of non-Hermitian (specifically, anti-Hermitian) Yukawa models
of interacting fermion and axion fields associated with their consistency as quantum field theoretic models, stemming, e.g., from microscopic models embedded in string theory~\cite{mp}. Apart from the anti-Hermitian Yukawa interactions, the models included anti-Hermitian anomaly terms, as well as interactions of the fermions with constant anti-Hermitian axial backgrounds. The underlying $\mathcal C PT$ symmetry \eqref{cpt} of these models guarantee~\cite{antilin,alex1,alex2,soto} the reality of the energy spectra, and thus the dynamical masses.

We have addressed the issue of dynamical mass generation for both fermions and axions, with the conclusion that in the absence of attractive four-fermion (Hermitian) interactions, only an axion mass can be generated dynamically, due to energetics. Par contrast, in the presence of sufficiently strong four-fermion attractive interactions, Yukawa-interaction-induced dynamical mass generation for {\it both} fermions and axions is possible. The role of the (sufficiently strong) four-fermion interactions, therefore, is that of a catalyst for the dynamical mass generation in this case. 
The anomaly terms, on the other hand, resist dynamical generation of fermion masses. 
The (constant) axial background assists the Yukawa-coupling-induced mass generation, in the presence of strong four-fermion interactions, in the sense that the larger the value of the background field, the larger the magnitude of the anti-Hermitian-Yukawa-interaction-induced dynamical mass.

In the context of microscopic string theories, the origin of the non-Hermitian background can be traced back to the non-Hermitian anomalous interactions of the KR gravitational axion of the massless bosonic string multiplet in some formulations of the theory ( see ``{\bf Scheme II}'', \eqref{sea6II},  in section \ref{sec:model}). It would be interesting to investigate further the consequences of such anti-Hermitian  axial backgrounds for Leptogenesis, as per the scenario of \cite{decesare,boss}. In general, non-Hermitian Yukawa interactions, of the type discussed here, may play a r\^ole in neutrino physics, including neutrino oscillations, as well as non-Hermitian extensions of the standard model. 

We conclude, by mentioning the recent work of \cite{cherno}, according to which anti-Hermitian Yukawa interactions, of the type discussed in  \cite{alex1,alex2} and reviewed here, may also arise as a result of linearising Hermitian four-fermion interactions, of Nambu-Jona-Lasinio type~\cite{NJL}, using appropriate complex auxiliary (pseudo)scalar fields. According to such an analysis, one has {\it spontaneous}  appearance of {\it non-Hermiticity} in these systems.  In addition, four-fermion interactions of Nambu-Jona-Lasinio type with {\it complex couplings}, whose imaginary part represents dissipation, have been discussed in \cite{kana}, with the conclusion that the imaginary coupling tends to enhance chiral-symmetry breaking up to a given threshold, above which the symmetry is restored. Such results could be testable in systems of ultracold atomic gases, or other materials that are characterised by Dirac-(Majorana-, or Weyl-)fermion-like excitations~\cite{diracmaterial,Wilc,weylmaterial}. 

As an outlook, we mention that, from a pure field-theoretic view point, the non-Hermitian models we have discussed here should be studied in a more rigorous way as far as their path-integral quantization is concerned. To this end,  we 
should combine our SD treatment for mass generation with an appropriate extension to fermionic models of the methods developed in \cite{qft5} for defining properly the path-integral measure of pseudoscalar fields in PT-symmetric pseudoscalar field theories, leading to their  renormalization. 
 Such a combined study might lead to a renormalization-group improved SD analysis for our models.

Moreover, an understanding of the concept of anomalies in such non-Hermitian quantum field theories is pressing. When, and if, this is accomplished, one could obtain a geometric understanding of the anti-Hermitian anomalous interactions that we focus our attention on in this review, and in \cite{soto}. A fully microscopic understanding of such terms in the context of UV-complete theories of quantum gravity, such as string/brane theory, is also desirable, but this might be a rather long shot. 

Another potentially interesting topic is the application of the non-Hermitian anomalous Yukawa models in the presence of (approximately constant) non-Hermitian axial backgrounds to the generation of a Lepton-number asymmetry in the early Universe (Leptogenesis), by extending appropriately the corresponding scenarios of the Hermitian case~\cite{decesare,boss}, provided, of course, that such an extension is possible. The dynamical fermion-mass generation in such non-Hermitian models, studied in \cite{alex1,soto} and reviewed above, when applied to Majorana neutrinos, might lead to Lepton-number asymmetries due to the asymmetric decays of such massive fields to 
standard-model leptons and antileptons in the presence of the non-Hermitian axial backgrounds. We hope to be able to come back to a study of some of these important issues in the near future.

We close this mini review with an optimistic note.  We believe that, although several aspects of non-Hermitian quantum field theories embeddable in a generalised-PT($\mathcal C$PT) framework are yet to be understood, nonetheless the rapidly-growing interest in such theories is a strong  indicator that soon they might found themselves  at comparable levels of rigour and physical applicability with their Hermitian counterparts. In this respect, it would be interesting to see whether there are fundamental aspects of particle physics models, perhaps in the dark sector,  that could be explained with such non-Hermitian interactions, which otherwise would remain a mystery. Such an example constituted the topic of this review,  namely the novel mechanisms involving anti-Hermitian Yukawa interactions for dynamical generation of  masses of right-handed neutrinos and axions, both of which can play the r\^ole of dark matter candidates in models beyond the standard model of particle physics. 

\ack

\medskip 

The author wishes to thank C.M. Bender, F. Correa  and A. Fring for the invitation to write this mini-review for the Proceedings of the virtual series on Pseudo-Hermitian Hamiltonians in Quantum Physics (XIX$<$vPHHQP$<$XX), which the author contributed to as a speaker of a seminar on the topics presented ihere. He also thanks Jean Alexandre, Sarben Sarkar  and Alex Soto for discussions and collaboration. 
This work is supported in part by the UK Science and Technology Facilities research
Council (STFC) under the research grant ST/T000759/1. The author acknowledges participation in the COST Association Action CA18108 {\it Quantum Gravity Phenomenology in the Multimessenger Approach (QG-MM)}. He also thanks  the Institute of Particle Physics (IFIC-CSIC) of the University of Valencia (Valencia, Spain ) for a 
scientific associateship ({\it Doctor Vinculado}).

%

\section*{References}
\medskip 

\bibliographystyle{iopart-num}

\bibliography{ptreview.bib}

\providecommand{\newblock}{}
\begin{thebibliography}{10}
\expandafter\ifx\csname url\endcsname\relax
  \def\url#1{{\tt #1}}\fi
\expandafter\ifx\csname urlprefix\endcsname\relax\def\urlprefix{URL }\fi
\providecommand{\eprint}[2][]{\url{#2}}

\bibitem{PTqm1}
Bender C~M and Boettcher S 1998 {\em Phys. Rev. Lett.\/} {\bf 80} 5243--5246
  (\textit{Preprint} \eprint{physics/9712001})

\bibitem{PTqm}
Bender C~M, Boettcher S and Meisinger P 1999 {\em J. Math. Phys.\/} {\bf 40}
  2201--2229 (\textit{Preprint} \eprint{quant-ph/9809072})

\bibitem{PT}
Bender C~M 2005 {\em Contemp. Phys.\/} {\bf 46} 277--292 (\textit{Preprint}
  \eprint{quant-ph/0501052})

\bibitem{PT2}
Bender C~M 2007 {\em Rept. Prog. Phys.\/} {\bf 70} 947 (\textit{Preprint}
  \eprint{hep-th/0703096})

\bibitem{PTappl}
Bender C~M 2015 {\em J. Phys. Conf. Ser.\/} {\bf 631} 012002

\bibitem{most2}
Mostafazadeh A 2002 {\em J. Math. Phys.\/} {\bf 43} 205--214 (\textit{Preprint}
  \eprint{math-ph/0107001})

\bibitem{most3}
Mostafazadeh A 2002 {\em J. Math. Phys.\/} {\bf 43} 2814--2816
  (\textit{Preprint} \eprint{math-ph/0110016})

\bibitem{most}
Mostafazadeh A 2010 {\em Int. J. Geom. Meth. Mod. Phys.\/} {\bf 7} 1191--1306
  (\textit{Preprint} \eprint{0810.5643})

\bibitem{maninnerPT}
Mannheim P~D 2018 {\em Phys. Rev. D\/} {\bf 97} 045001 (\textit{Preprint}
  \eprint{1708.01247})

\bibitem{antilin}
Mannheim P~D 2018 {\em J. Phys. A\/} {\bf 51} 315302 (\textit{Preprint}
  \eprint{1512.04915})

\bibitem{cherno2}
Chernodub M 2017 {\em J. Phys. A\/} {\bf 50} 385001 (\textit{Preprint}
  \eprint{1701.07426})

\bibitem{qft1a}
Bender C~M, Milton K~A and Savage V~M 2000 {\em Phys. Rev. D\/} {\bf 62} 085001
  (\textit{Preprint} \eprint{hep-th/9907045})

\bibitem{qft1b}
Bender C~M and Jones H 2000 {\em Found. Phys.\/} {\bf 30} 393--411

\bibitem{qft1c}
Bender C~M, Brody D~C and Jones H~F 2004 {\em Phys. Rev. D\/} {\bf 70} 025001
  [Erratum: Phys.Rev.D 71, 049901 (2005)] (\textit{Preprint}
  \eprint{hep-th/0402183})

\bibitem{qft1d}
Bender C~M, Brandt S~F, Chen J~H and Wang Q~h 2005 {\em Phys. Rev. D\/} {\bf
  71} 065010 (\textit{Preprint} \eprint{hep-th/0412316})

\bibitem{qft1e}
Bender C~M and Klevansky S 2010 {\em Phys. Rev. Lett.\/} {\bf 105} 031601
  (\textit{Preprint} \eprint{1002.3253})

\bibitem{qft2}
Bender C~M, Jones H and Rivers R 2005 {\em Phys. Lett. B\/} {\bf 625} 333--340
  (\textit{Preprint} \eprint{hep-th/0508105})

\bibitem{qft3}
Bender C~M, Branchina V and Messina E 2012 {\em Phys. Rev. D\/} {\bf 85} 085001
  (\textit{Preprint} \eprint{1201.1244})

\bibitem{qft4}
Bender C~M, Branchina V and Messina E 2013 {\em Phys. Rev. D\/} {\bf 87} 085029
  (\textit{Preprint} \eprint{1301.6207})

\bibitem{AB}
Alexandre J and Bender C~M 2015 {\em J. Phys. A\/} {\bf 48} 185403
  (\textit{Preprint} \eprint{1501.01232})

\bibitem{AMS}
Alexandre J, Millington P and Seynaeve D 2017 {\em Phys. Rev. D\/} {\bf 96}
  065027 (\textit{Preprint} \eprint{1707.01057})

\bibitem{qft5}
Bender C~M, Hassanpour N, Klevansky S and Sarkar S 2018 {\em Phys. Rev. D\/}
  {\bf 98} 125003 (\textit{Preprint} \eprint{1810.12479})

\bibitem{ptf}
Beygi A, Klevansky S and Bender C~M 2019 {\em Phys. Rev. A\/} {\bf 99} 062117
  (\textit{Preprint} \eprint{1904.00878})

\bibitem{neu1}
Alexandre J, Bender C~M and Millington P 2015 {\em JHEP\/} {\bf 11} 111
  (\textit{Preprint} \eprint{1509.01203})

\bibitem{neu2}
Alexandre J, Bender C~M and Millington P 2017 {\em J. Phys. Conf. Ser.\/} {\bf
  873} 012047 (\textit{Preprint} \eprint{1703.05251})

\bibitem{hg1}
Alexandre J, Ellis J, Millington P and Seynaeve D 2018 {\em Phys. Rev. D\/}
  {\bf 98} 045001 (\textit{Preprint} \eprint{1805.06380})

\bibitem{hg2}
Alexandre J, Ellis J, Millington P and Seynaeve D 2019 {\em Phys. Rev. D\/}
  {\bf 99} 075024 (\textit{Preprint} \eprint{1808.00944})

\bibitem{hg3}
Alexandre J, Ellis J, Millington P and Seynaeve D 2020 {\em Phys. Rev. D\/}
  {\bf 101} 035008 (\textit{Preprint} \eprint{1910.03985})

\bibitem{hg4}
Mannheim P~D 2019 {\em Phys. Rev. D\/} {\bf 99} 045006 (\textit{Preprint}
  \eprint{1808.00437})

\bibitem{hg5}
Fring A and Taira T 2020 {\em Nucl. Phys. B\/} {\bf 950} 114834
  (\textit{Preprint} \eprint{1906.05738})

\bibitem{hg5b}
Fring A and Taira T 2020 {\em Phys. Rev. D\/} {\bf 101} 045014
  (\textit{Preprint} \eprint{1911.01405})

\bibitem{hg6}
Fring A and Taira T 2020  (\textit{Preprint} \eprint{2004.00723})

\bibitem{ptdisc}
Alexandre J, Ellis J and Millington P 2020  (\textit{Preprint}
  \eprint{2006.06656})

\bibitem{ptsusy}
Alexandre J, Ellis J and Millington P 2020 {\em Phys. Rev. D\/} {\bf 101}
  085015 (\textit{Preprint} \eprint{2001.11996})

\bibitem{chernochiral}
Chernodub M~N and Cortijo A 2020 {\em Symmetry\/} {\bf 12} 761
  (\textit{Preprint} \eprint{1901.06167})

\bibitem{liouv}
Bender C~M, Hook D~W, Mavromatos N~E and Sarkar S 2014 {\em Phys. Rev. Lett.\/}
  {\bf 113} 231605 (\textit{Preprint} \eprint{1408.2432})

\bibitem{stab}
Bender C~M, Hook D~W, Mavromatos N~E and Sarkar S 2016 {\em J. Phys. A\/} {\bf
  49} 45LT01 (\textit{Preprint} \eprint{1506.01970})

\bibitem{ptmon}
Fring A and Taira T 2020 {\em Phys. Lett. B\/} {\bf 807} 135583
  (\textit{Preprint} \eprint{2006.02718})

\bibitem{bpssol}
Fring A and Taira T 2020  (\textit{Preprint} \eprint{2007.15425})

\bibitem{alex1}
Alexandre J and Mavromatos N~E 2020 {\em Phys. Lett. B\/} {\bf 807} 135562
  (\textit{Preprint} \eprint{2004.03699})

\bibitem{alex2}
Alexandre J, Mavromatos N~E and Soto A 2020  (\textit{Preprint}
  \eprint{2004.04611})

\bibitem{soto}
Mavromatos N~E and Soto A 2020  (\textit{Preprint} \eprint{2006.13616})

\bibitem{mp}
Mavromatos N~E and Pilaftsis A 2012 {\em Phys. Rev. D\/} {\bf 86} 124038
  (\textit{Preprint} \eprint{1209.6387})

\bibitem{str1}
Green M~B, Schwarz J~H and Witten E 2012 {\em {Superstring Theory Vol. 1}:
  {25th Anniversary Edition}\/} Cambridge Monographs on Mathematical Physics
  (Cambridge University Press) ISBN 978-1-139-53477-2, 978-1-107-02911-8

\bibitem{str2}
Green M~B, Schwarz J~H and Witten E 2012 {\em {Superstring Theory Vol. 2}:
  {25th Anniversary Edition}\/} Cambridge Monographs on Mathematical Physics
  (Cambridge University Press) ISBN 978-1-139-53478-9, 978-1-107-02913-2

\bibitem{pol1}
Polchinski J 2007 {\em {String theory. Vol. 1: An introduction to the bosonic
  string}\/} Cambridge Monographs on Mathematical Physics (Cambridge University
  Press) ISBN 978-0-511-25227-3, 978-0-521-67227-6, 978-0-521-63303-1

\bibitem{pol2}
Polchinski J 2007 {\em {String theory. Vol. 2: Superstring theory and
  beyond}\/} Cambridge Monographs on Mathematical Physics (Cambridge University
  Press) ISBN 978-0-511-25228-0, 978-0-521-63304-8, 978-0-521-67228-3

\bibitem{eguchi}
Eguchi T, Gilkey P~B and Hanson A~J 1980 {\em Phys. Rept.\/} {\bf 66} 213

\bibitem{alvarez}
Alvarez-Gaume L and Witten E 1984 {\em Nucl. Phys. B\/} {\bf 234} 269

\bibitem{weinberg}
Weinberg S 2013 {\em {The quantum theory of fields. Vol. 2: Modern
  applications}\/} (Cambridge University Press) ISBN 978-1-139-63247-8,
  978-0-521-67054-8, 978-0-521-55002-4

\bibitem{hehl}
Hehl F, Von Der~Heyde P, Kerlick G and Nester J 1976 {\em Rev. Mod. Phys.\/}
  {\bf 48} 393--416

\bibitem{hull}
Hull C 1986 {\em Phys. Lett. B\/} {\bf 167} 51--55

\bibitem{mavindex}
Mavromatos N 1988 {\em J. Phys. A\/} {\bf 21} 2279

\bibitem{kaloper}
Duncan M~J, Kaloper N and Olive K~A 1992 {\em Nucl. Phys. B\/} {\bf 387}
  215--235

\bibitem{witten}
Svrcek P and Witten E 2006 {\em JHEP\/} {\bf 06} 051 (\textit{Preprint}
  \eprint{hep-th/0605206})

\bibitem{boss}
Bossingham T, Mavromatos N~E and Sarkar S 2018 {\em Eur. Phys. J. C\/} {\bf 78}
  113 (\textit{Preprint} \eprint{1712.03312})

\bibitem{anomalies}
Basilakos S, Mavromatos N~E and Sol\`a~Peracaula J 2020 {\em Phys. Rev. D\/}
  {\bf 101} 045001 (\textit{Preprint} \eprint{1907.04890})

\bibitem{strom}
Giddings S~B and Strominger A 1988 {\em Nucl. Phys. B\/} {\bf 306} 890--907

\bibitem{decesare}
de~Cesare M, Mavromatos N~E and Sarkar S 2015 {\em Eur. Phys. J. C\/} {\bf 75}
  514 (\textit{Preprint} \eprint{1412.7077})

\bibitem{basilakos}
Basilakos S, Mavromatos N~E and Sol\`a~Peracaula J 2020 {\em Phys. Lett. B\/}
  {\bf 803} 135342 (\textit{Preprint} \eprint{2001.03465})

\bibitem{shapiro}
de~Berredo-Peixoto G, Freidel L, Shapiro I and de~Souza C 2012 {\em JCAP\/}
  {\bf 06} 017 (\textit{Preprint} \eprint{1201.5423})

\bibitem{Kim1}
Kim J~E 1987 {\em Phys. Rept.\/} {\bf 150} 1--177

\bibitem{kim}
Kim J~E and Carosi G 2010 {\em Rev. Mod. Phys.\/} {\bf 82} 557--602 [Erratum:
  Rev.Mod.Phys. 91, 049902 (2019)] (\textit{Preprint} \eprint{0807.3125})

\bibitem{astro1}
Raffelt G~G 2008 {\em Lect. Notes Phys.\/} {\bf 741} 51--71 (\textit{Preprint}
  \eprint{hep-ph/0611350})

\bibitem{astro2}
Arvanitaki A and Dubovsky S 2011 {\em Phys. Rev. D\/} {\bf 83} 044026
  (\textit{Preprint} \eprint{1004.3558})

\bibitem{astro3}
Arvanitaki A, Baryakhtar M and Huang X 2015 {\em Phys. Rev. D\/} {\bf 91}
  084011 (\textit{Preprint} \eprint{1411.2263})

\bibitem{marsh}
Marsh D~J~E 2016 {\em Phys. Rept.\/} {\bf 643} 1--79 (\textit{Preprint}
  \eprint{1510.07633})

\bibitem{astro4}
Arvanitaki A, Baryakhtar M, Dimopoulos S, Dubovsky S and Lasenby R 2017 {\em
  Phys. Rev. D\/} {\bf 95} 043001 (\textit{Preprint} \eprint{1604.03958})

\bibitem{latticeqcd}
Grilli~di Cortona G, Hardy E, Pardo~Vega J and Villadoro G 2016 {\em JHEP\/}
  {\bf 01} 034 (\textit{Preprint} \eprint{1511.02867})

\bibitem{arvanitaki}
Arvanitaki A, Dimopoulos S, Dubovsky S, Kaloper N and March-Russell J 2010 {\em
  Phys. Rev. D\/} {\bf 81} 123530 (\textit{Preprint} \eprint{0905.4720})

\bibitem{shiu1}
Shiu G, Staessens W and Ye F 2015 {\em Phys. Rev. Lett.\/} {\bf 115} 181601
  (\textit{Preprint} \eprint{1503.01015})

\bibitem{shiu2}
Shiu G, Staessens W and Ye F 2015 {\em JHEP\/} {\bf 06} 026 (\textit{Preprint}
  \eprint{1503.02965})

\bibitem{ib1}
Ibanez L and Uranga A 2007 {\em JHEP\/} {\bf 03} 052 (\textit{Preprint}
  \eprint{hep-th/0609213})

\bibitem{ib2}
Ibanez L, Schellekens A and Uranga A 2007 {\em JHEP\/} {\bf 06} 011
  (\textit{Preprint} \eprint{0704.1079})

\bibitem{cve1}
Blumenhagen R, Cvetic M and Weigand T 2007 {\em Nucl. Phys. B\/} {\bf 771}
  113--142 (\textit{Preprint} \eprint{hep-th/0609191})

\bibitem{cve2}
Cvetic M, Richter R and Weigand T 2007 {\em Phys. Rev. D\/} {\bf 76} 086002
  (\textit{Preprint} \eprint{hep-th/0703028})

\bibitem{donogh}
Donoghue J~F 1994 {\em Phys. Rev. D\/} {\bf 50} 3874--3888 (\textit{Preprint}
  \eprint{gr-qc/9405057})

\bibitem{benakli}
Antoniadis I, Benakli K and Laugier A 2001 {\em JHEP\/} {\bf 05} 044
  (\textit{Preprint} \eprint{hep-th/0011281})

\bibitem{NJL}
Nambu Y and Jona-Lasinio G 1961 {\em Phys. Rev.\/} {\bf 122} 345--358

\bibitem{VW}
Vafa C and Witten E 1984 {\em Nucl. Phys. B\/} {\bf 234} 173--188

\bibitem{bashir}
Bashir A and Diaz-Cruz J 1999 {\em J. Phys. G\/} {\bf 25} 1797--1805
  (\textit{Preprint} \eprint{hep-ph/9906360})

\bibitem{NJLnh}
Felski A, Beygi A and Klevansky S 2020 {\em Phys. Rev. D\/} {\bf 101} 116001
  (\textit{Preprint} \eprint{2004.04011})

\bibitem{cherno}
Chernodub M, Cortijo A and Ruggieri M 2020  (\textit{Preprint}
  \eprint{2008.11629})

\bibitem{kana}
Kanazawa T 2020  (\textit{Preprint} \eprint{2009.13363})

\bibitem{diracmaterial}
Vafek O and Vishwanath A 2014 {\em Ann. Rev. Condensed Matter Phys.\/} {\bf 5}
  83--112 (\textit{Preprint} \eprint{1306.2272})

\bibitem{Wilc}
Wilczek F 2014  (\textit{Preprint} \eprint{1404.0637})

\bibitem{weylmaterial}
Armitage N, Mele E and Vishwanath A 2018 {\em Rev. Mod. Phys.\/} {\bf 90}
  015001 (\textit{Preprint} \eprint{1705.01111})

\end{thebibliography}

\end{document}